\documentclass[preprint,12pt]{elsarticle}

\usepackage{graphics}
\usepackage{graphicx}
\usepackage[utf8]{inputenc}
\usepackage{epsfig}
\usepackage{booktabs}
\usepackage{amssymb}
\usepackage{amsmath}
\usepackage{tabularx}
\usepackage{lineno}
\usepackage{capt-of}
\usepackage{listings}
\usepackage{xcolor}

\definecolor{codegreen}{rgb}{0,0.6,0}
\definecolor{codegray}{rgb}{0.5,0.5,0.5}
\definecolor{codepurple}{rgb}{0.58,0,0.82}
\definecolor{backcolour}{rgb}{0.95,0.95,0.92}

\lstdefinestyle{mystyle}{
    backgroundcolor=\color{backcolour},   
    commentstyle=\color{codegreen},
    keywordstyle=\color{magenta},
    numberstyle=\tiny\color{codegray},
    stringstyle=\color{codepurple},
    basicstyle=\ttfamily\footnotesize,
    breakatwhitespace=false,         
    breaklines=true,                 
    captionpos=b,                    
    keepspaces=true,                 
    numbers=left,                    
    numbersep=5pt,                  
    showspaces=false,                
    showstringspaces=false,
    showtabs=false,                  
    tabsize=2
}

\usepackage{hyperref}
\hypersetup{
    colorlinks=true,
    linkcolor=blue,
    filecolor=magenta,      
    urlcolor=cyan,
    citecolor=red,
}

\usepackage{caption}
\usepackage{subcaption}

\journal{Computer Physics Communications}

\begin{document}

\begin{frontmatter}


\title{\texttt{MF-toolkit}: A High-Performance Python Library for Multifractal Analysis with Automated Crossover Detection, Source Identification and Application to Gravitational Waves Data. }
\cortext[cor1]{Corresponding author}
\author[1,2]{Mendez, N.\corref{cor1}}
\ead{nmendez@frh.utn.edu.ar}
\author[3]{Mariani, M.}
\author[3]{Beccar-Varela, M.}
\author[4]{Tweneboah, O.}
\author[2,5]{Jaroszewicz, S.}

\address[1]{Instituto Sabato, Universidad Nacional de San Martín, 1650, Buenos Aires, Argentina}
\address[2] {Facultad Regional Haedo, Universidad Tecnológica Nacional, 1706, Buenos Aires, Argentina}
\address[3]{Department of Mathematical Science, UTEP, 79968, El Paso, USA}
\address[4]{Data Science Program. Ramapo College of New Jersey, New Jersey, 07430, USA.}
\address[5]{Departamento de Materia Condensada, Comisión Nacional de Energía Atómica, 1650, Buenos Aires, Argentina}

\begin{abstract}
Multifractal Detrended Fluctuation Analysis (MFDFA) is a powerful and widely used technique for characterizing the scaling properties and long-range correlations of complex time series. However, its application often involves significant practical challenges, such as the subjective identification of scaling regions (crossovers) and the disambiguation of the physical origins of multifractality. We introduce MF-toolkit, a high-performance, parallelized Python library designed to address these challenges. It integrates three key innovations: (1) fully automatic crossover detection algorithms (CDV-A and SPIC), which remove operator bias and enhance reproducibility; (2) a built-in implementation of the Iterative Amplitude Adjusted Fourier Transform (IAAFT) for generating surrogate data, enabling the robust identification of the source of multifractality; and (3) a comprehensive suite for generating synthetic time series for rigorous validation. We demonstrate the rigor and utility of MF-toolkit through its application to characterize the multifractal properties of non-stationary noise in gravitational wave (LIGO) data. The MF-toolkit library offers a robust, efficient, and user-friendly tool for advanced time series analysis, facilitating more rigorous and reproducible research across physics and other data-intensive fields.
\end{abstract}


\end{frontmatter}
\noindent \textbf{PROGRAM SUMMARY/NEW VERSION PROGRAM SUMMARY}

\noindent
{\em Program Title:}  MF-toolkit                                        \\
{\em CPC Library link to program files:} (to be added by Technical Editor) \\
{\em Developer's repository link:} (if available) https://github.com/NahueMendez/mf-toolkit \\
{\em Licensing provisions:} MIT  \\
{\em Programming language:}   Python                                \\
{\em Supplementary material:} https://app.readthedocs.org/projects/mf-toolkit/                                 \\
{\em Journal reference of previous version:}*                  \\
{\em Does the new version supersede the previous version?:}*   \\
{\em Reasons for the new version:*}\\
{\em Summary of revisions:}*\\
{\em Nature of problem(approx. 50-250 words):} Applying multifractal detrended fluctuation analysis to a time series is a widely used process in complex system physics but the time computing and the resources needed strongly depends on the number of samples of the input signal. The method in its multifractal extension require to calculate the fluctuation function for every scale size $s$ and every moment $q$. This issue escalates when it is needed to analyze a group of time series to asses statistically its multifractals parameters.\\
{\em Solution method(approx. 50-250 words):} We address the bottleneck of MFDFA by computing the fluctuation functions using CPU-based plane parallelization and Numba's just-in-time compilation taking advantage of the independency of the calculations. This drastically reduces time computing, and it can be used for processing large datasets.\\
{\em Additional comments including restrictions and unusual features (approx. 50-250 words):}\\
   \\
   

\section{Introduction}
\label{sec:intro}
Time series from complex systems, ranging from fluid turbulence and financial markets\cite{Jaroszewicz2005} to biological signals \cite{Peng1994,jaroszewicz_1,Mendez04032026}, climate \cite{Maruyama2011,JAROSZEWICZ2024106161} and geophysical data \cite{Varo_5, Varo_7, Ribeiro2014}, often exhibit scale-invariant structures characteristic of fractals \cite{Mandelbrot1968, Mariani2020a, Mariani2020b}. Techniques like Detrended Fluctuation Analysis (DFA) have emerged as a robust and standard method to quantify this structure with a scaling exponent called Hurst exponent ($H$). A generalization of the previous method named Multifractal Detrended Fluctuation Analysis (MFDFA) has been developed that goes beyond a single scaling exponent to describe a full spectrum of singularities $f(\alpha)$ by studying the moments of the fluctuations of the series \cite{Kantelhardt2002, Ihlen2012, Kantelhardt2008}.

\vspace{5mm}

Despite its undeniable power, the practical application of MFDFA is fraught with challenges that can compromise the objectivity and reproducibility of results. The most significant challenges are the presence of crossovers in the fluctuation functions $F_q(s)$, which mislead the linear fit, and the subjective selection of optimal scaling ranges and moments $(q)$ \cite{Gulich2014}. These issues necessitate a comprehensive tool that provides automated and statistically robust checks for scaling regimes, model fitting, and compliance with theoretical multifractal properties (such as the concavity of $f(\alpha)$).

\vspace{5mm}

In addition, the library was designed to be applied to a large number of datasets, optimizing the resources and expediting the results. This was done using a CPU-based parallelization that computes the fluctuation functions faster for each moment $q$ by taking advantage of their independence. 
\\
Characterizing noise in cutting-edge physics experiments presents a formidable computational and analytical challenge. For example, in gravitational wave (GW) astronomy, strain data from detectors such as LIGO are dominated by intricate dynamics and colored noise that often exhibits complex multifractal correlations \cite{de2018multifractal,oliveira2025multifractal,chernogor2017multi,cavaglia2022characterization}. Distinguishing a transient astrophysical signal such as a black hole coalescence from this multifractal background requires objective, fast, and rigorous time series analysis to identify the source of multifractality. The application of MFDFA to large GW datasets therefore not only requires high-performance implementation to handle long series, but also the inclusion of automatic tools that eliminate subjectivity in the selection of the scaling range and validate the physical source of the observed correlations.

\vspace{5mm}

While several MFDFA implementations exist, including a notable Python library that focused on efficiency and common extensions \cite{RydinGorjo2022}, a comprehensive tool that systematically addresses the aforementioned challenges has been lacking. To fill this gap, we developed \texttt{MF-toolkit}, a new high-performance Python library. Our implementation is parallelized for speed and introduces a suite of advanced features designed for rigorous and automated multifractal analysis. The key contributions of our library are:
\begin{itemize}
    \item An integrated, automatic crossover detection module using specialized algorithms such as \textit{CDV-A} \cite{Moreno-Pulido2025} and \textit{SPIC} \cite{Ge2013}.
    
    \item A built-in tool for analyzing the origin of multifractality using the \textit{Iterative Amplitude Adjusted Fourier Transform (IAAFT)} surrogate method \cite{Schreiber2000, Theiler1992}.
    \item A powerful class for generating synthetic time series with a controlled (or "pure") source of multifractality, creating an orthogonal testbed for validation.
\end{itemize}

This paper introduces the \texttt{MF-toolkit} library. Section \ref{sec:theory} reviews the theoretical background of MFDFA, the origins of multifractality, and the crossover phenomenon. Section \ref{sec:results} details the implementation of our library's novel features and demonstrates the library's capabilities through applications to both synthetic and real-world data. Section \ref{sec:performance} discusses performance and availability, and Section \ref{sec:conclusion} provides concluding remarks.

\section{Theoretical Framework}
\label{sec:theory}

\subsection{Multifractal Detrended Fluctuation Analysis (MFDFA)}
The MFDFA algorithm, as detailed by Kantelhardt et al. \cite{Kantelhardt2002}, aims to determine the scaling properties of a time series $x_i$ for $i=1, \dots, N$. The procedure can be summarized as follows:

\begin{enumerate}
    \item The mean-subtracted cumulative sum (the "profile") $Y_i$ is calculated:
    $$ Y_{i} = \sum_{k=1}^{i} (x_{k} - \langle x \rangle) $$
    \item The profile $Y_i$ is divided into $N_s = \text{int}(N/s)$ non-overlapping segments of length $s$. To utilize the full series, this is repeated from the end of the series, yielding $2N_s$ segments in total.
    \item For each segment $\nu$, a local trend is estimated by fitting a polynomial of order $m$, denoted $y_{\nu,i}$. The variance of the residual is then calculated:
    $$ F^2(\nu, s) = \frac{1}{s}\sum_{i=1}^{s} [Y_{(\nu-1)s+i} - y_{\nu,i}]^2 $$

    In our code we have integrated a warning to the user that notes of a not good linear fit in $F_q(s)$ for the range selected. This warning is based in the assessment of the regression parameters, e.g $R^2>0.85$ and p-value$<0.05$, and then an evaluation of the normality of the residuals using the Shapiro test and a p-value$<0.05$. If none of this conditions is verified, the function warns the user of a poor fit. 
    
    \item The $q$-th order fluctuation function, $F_q(s)$, is obtained by averaging over all segments:
    \begin{equation} 
    \label{eq:q-Fluctuations}
    F_q(s) = \left\{ \frac{1}{2N_s} \sum_{\nu=1}^{2N_s} [F^2(\nu, s)]^{q/2} \right\}^{1/q} 
    \end{equation}
    For $q=0$, a logarithmic average is used. If the time series exhibits fractal scaling, $F_q(s)$ follows a power law:
    \begin{equation}
    \label{eq:Hurst_exp}
       F_q(s) \sim s^{h(q)} 
    \end{equation}

    where $h(q)$ is the generalized Hurst exponent. For monofractal series, $h(q)$ is constant and equal to the Hurst exponent $H$. For multifractal series, $h(q)$ is a non-constant function of $q$.
    \item The multifractal scaling exponent $\tau(q)$ and the singularity spectrum $f(\alpha)$ are derived from $h(q)$ via:
    $$ \tau(q) = qh(q) - 1 $$
    $$ \alpha = \frac{d\tau(q)}{dq}, \quad f(\alpha) = q\alpha - \tau(q) $$
    The width of the singularity spectrum, $\omega = \alpha_{max} - \alpha_{min}$, quantifies the degree of multifractality. Additionally, other parameters can be used as the position of the spectral maximum, $\alpha_0 = \alpha(q=0)$, which characterizes the average correlation structure of the signal (anti-persistent for $\alpha_0<0.5$, persistent for $\alpha_0>0.5$); and the spectral asymmetry, $\alpha_{asy}=\frac{(\alpha_{0}-\alpha_{min})-(\alpha_{max}-\alpha_{0})}{\omega}$, which describes the balance of fluctuations driving the complexity.

    \vspace{3mm}

    In our software implementation, we enforce automated validation checks on these final derived quantities to ensure they adhere to the foundational properties of the multifractal formalism. Specifically, the toolkit verifies that the singularity spectrum satisfies the strict topological bound $0 \le f(\alpha) \le 1$. Because $f(\alpha)$ mathematically represents the Hausdorff dimension of the fractal subset characterized by a specific singularity strength $\alpha$, it is a topological impossibility for this dimension to exceed the dimension of the underlying one-dimensional time series support \cite{Kantelhardt2002}. Furthermore, the algorithm strictly checks that the $f(\alpha)$ curve exhibits downward concavity (a single-humped shape). Since $f(\alpha)$ is derived from the mass exponent $\tau(q)$ via a Legendre transform, it must inherently maintain negative concavity. Integrating these theoretically grounded constraints directly into the software provides a robust layer of quality control, preventing the misinterpretation of numerical artifacts—such as finite-size effects or insufficient scaling ranges—as genuine multifractality.
\end{enumerate}

\subsection{Iterated Amplitude Adjusted Fourier Transform (IAAFT)}

As stated by Schreiber et al.\cite{Schreiber2000} the Iterated Amplitude Adjusted Fourier Transform (IAAFT) method is designed for generating surrogate time series that preserves the power spectrum and statistical distribution of amplitudes of the original series. 
Let $X(f)$ be the Discrete Fourier Transform of the original time series $x(t)$:
$$X(f)=\sum_{n=0}^{N-1}x_ne^{-2\pi ft/N}$$
Therefore, the power spectrum is defined as the square oh the Fourier Transform magnitudes:
$$S_x(f)=|X(f)|^2$$
To initialize the algorithm it is needed to generate a white noise signal $z_t^0$ of the same length $N$ as the original. Then the original serie $x_t$ must be sorted in ascending order, as well as $z_t^{0}$. Finally it is needed to map or readjust the $z_t^0$ values to ordered values of $x_t$ preserving the rank order of $z_t^0$. 

This guarantees that the surrogate time series will have the same amplitude distribution as the original. 

\vspace*{2mm}

Once the initialization is finished the method consists in the iteration $k$ times of the following steps:
\begin{enumerate}
    \item Calculate the Fourier Transform of the candidate surrogate time series $z_t^k$:
    $$Z(f)^{(k)} = |Z(f)^{(k)}|e^{i\phi k(f)}$$
    \item Replace the magnitudes of the FT with the magnitudes of the original series, generating a new function that we can name $Z'(f)$:
    $$Z'(f) = |X(f)|e^{i\phi k(f)}$$
    \item Calculate the inverse Fourier Transform to obtain a new time series $z'_t$ with the identical power spectrum of the original series.
    $$z_t^{'(k)}=\frac{1}{N}\sum_{f=0}^{N-1}Z'(f)e^{i2\pi ft/N}$$
    \item Readjust the amplitudes that could be altered, by applying the original amplitudes of $x_t$ preserving the rank order of $z_t^{'(k)}$. The resulting series is called $z_t^{(k+1)}$ and is ready for a new iteration.
\end{enumerate}

\vspace{2mm}

The convergence criterion is usually associated with the stability of the power spectrum, such as the squared mean difference between $S_x(f)$ and $S_z(f)$, or alternatively a maximum number of iterations $k_{\text{max}}$ can be defined.

\subsection{Crossover Detection based on Variance of slopes differences Algorithm (CDV-A)}

Equation \ref{eq:Hurst_exp} assumes a single scaling regime for the fluctuation functions given by \ref{eq:q-Fluctuations}. However, many real-world systems exhibit different correlation properties at different time scales. This results in a "break" in the power-law behavior of $F_q(s)$, known as a \textit{crossover}. The log-log plot of $F_q(s)$ vs. $s$ will appear as two or more linear segments with different slopes. Failing to identify these crossovers and fitting a single line across them leads to erroneous estimation of $h(q)$ and a distorted multifractal spectrum. Moreno-Pulido et al. \cite{Moreno-Pulido2025} had developed an algorithm for detecting crossovers based on systematically identifying the scale at which the difference in local scaling exponents is maximized. 

\vspace{2mm}

The method can be summarized in the following steps:

\begin{enumerate}
    \item Compute a matrix of slope differences ($M$): For each possible crossover index $i$ (from $s_{min}$ to $s_{max}$) and for each possible number of fitting points $N_{fit}$, the algorithm calculates the slopes of the log-log plot of $F_q(s)$ to the left ($m_{left}$) and right ($m_{right}$) of $i$. The absolute difference $|m_{left} - m_{right}|$ is stored. There are two options for obtaining a single matrix of slope differences. The first is to repeat the calculations for all $q$ moments and average them to form a single matrix $M$, where rows correspond to $N_{fit}$ and columns to the scale index $i$. The second possibility is to compute the matrix only for $q=2$ and assess the crossovers in that case. 
    
    \item Sub-matrix selection via variances: Real fluctuation functions are noisy. This noise particularly affects fits with few points (low $N_{fit}$). The CDV-A method robustly mitigates this by analyzing the variance of the rows and columns of $M$.
        \begin{itemize}
            \item The variances of the rows are used to find a cutoff $N_{cut}$, discarding the noisy region of low $N_{fit}$.
            \item The variances of the columns of the remaining matrix are then calculated. An ideal crossover at index $i_x$ corresponds to a column with near-zero variance (as the slope difference is constant regardless of $N_{fit}$), surrounded by high-variance columns. This region is identified as the "valley".
        \end{itemize}
    In this step, the algorithm selects a sub-matrix $M_{sel}$ corresponding to the rows above $N_{cut}$ and the columns within the valley
    \item Crossover identification via norm $L_1$. The final crossover location is determined as the column in $M_{sel}$ that has the maximum $L_1$-norm (sum of absolute values). This provides a single, objective crossover index $i_x$.
\end{enumerate}

This procedure transforms the subjective task of visual inspection into a deterministic and reproducible algorithm integrated directly into the analysis workflow.

\subsection{Sequential Permutation for Identifying Crossovers}
This method was developed by Ge et. al in \cite{Ge2013} and its based on an iterative hypothesis testing procedure to find the correct number of crossover points and their locations.

\vspace{2mm}

The key steps of the algorithm are listed as follows:
\begin{enumerate}
    \item Define a general regression model for $k$ crossovers, with $\chi=log(s)$ and $\Gamma=log(F_q)$ for a given q:
    $$\hat{\Gamma}=\beta_{10}+\beta_{11}\chi+\sum_{k=1}^{K}(\beta_{k+1,1}-\beta_{k,1})(\chi-\tau_k)^+$$

    Being $\beta_{10}$ and $\beta_{11}$ the intercept and the slope for the first regime, meanwhile the rest terms depicts the fluctuations in the kth subregion when $\chi > \tau_k$, and $(\chi-\tau_k)^+=0$ when $\chi \leq t_k$. The $t_k$ values represent the crossover time scales to be determined.
    \item Formulate hypothesis test: An iterative statistical inference procedure is used to determine the optimal number of crossover points. This involves testing a null hypothesis ($H_0$) of $k_0$ crossover points against an alternative hypothesis ($H_1$) of $k_1$ crossover points, where $k_0 < k_1$. The procedure continues until the number of crossover points is statistically determined.
    \item Fit the model using a grid-search method: To estimate the parameters of the regression model and find the best fit, a grid-search method is employed. This method minimizes the residual sum of squared error (SSE) by testing all possible combinations of crossover points. It also provides a way to calculate the confidence interval for the crossover points.
    $$SSE=\sum_{i=0}^{n}(\Gamma_i-\hat{\Gamma^{(k)}_i})^2$$

    Being $\epsilon^{(k)}_i = (\Gamma_i-\hat{\Gamma^{(k)}_i})$ the residuals.
    
    \item Use a Monte Carlo method called permutation test: To determine if the difference between the hypothesized models is statistically significant, a test statistic, $T(\Gamma)$, is calculated based on the ratio of the residual sums of squares of the null and alternative models. 
    Since the exact distribution of $T(\Gamma)$ is unknown, a Monte Carlo method, specifically a permutation test, is used to approximate its distribution.
    $$T(\Gamma)=\frac{|\hat{\epsilon^{(k_0)}(\Gamma)}|'|\hat{\epsilon^{(k_0)}(\Gamma)}|}{|\hat{\epsilon^{(k_1)}(\Gamma)}|'|\hat{\epsilon^{(k_1)}(\Gamma)}|}$$
    \item Calculate the p-value: The permutation test works by repeatedly permuting the residuals of the null model and adding them back to the null-modeled means to create a set of permuted data. For each permuted dataset, the test statistic $T(\Gamma_{(a)})$ is calculated with $a$ from $1$ to the maximum number of permutations $N_p$ . The p-value is then determined by comparing the observed test statistic $T(\Gamma)$ from the original data to the distribution of $T(\Gamma_{(a)})$ values from the permutations. This p-value is used to decide whether to accept or reject the null hypothesis, thereby statistically determining the number of crossover points and their location.
\end{enumerate}

\subsection{Validation of Synthetic Data Properties}
\subsubsection{Monofractal Serie}
The generation of a stationary fractional Gaussian noise (fGn) series, which integrates to a fractional Brownian motion (fBm), can be achieved using the Davies-Harte method \cite{daviesharte87}. This algorithm leverages the properties of circulating matrices and the Fast Fourier Transform (FFT) to efficiently generate a series with a given Hurst exponent $H$. For a desired fGn series length of $N$, the algorithm proceeds as follows:

\begin{enumerate}
\item Algorithm Initialization: The method operates on an extended circulating grid of size $M = 2(N-1)$. A vector $\mathbf{c}$ of length $M$, which forms the first row of a circulating covariance matrix, is constructed using the auto-covariance function $\gamma_H(k)$ of the fGn process.

\item Eigenvalue Computation via FFT: The eigenvalues $\boldsymbol{\lambda}$ of the circulating covariance matrix are obtained by computing the FFT of the vector $\mathbf{c}$. The method relies on the fact that these eigenvalues correspond to the power spectrum of the fGn process on the extended grid.

\item Frequency Domain Generation: A vector $\mathbf{Z}$ of $M$ independent complex Gaussian random variables is generated, with independent real and imaginary parts. A frequency-domain vector $\mathbf{Y}$ is then constructed by scaling these complex numbers with the square root of the eigenvalues: $\mathbf{Y} = \sqrt{\frac{\boldsymbol{\lambda}}{2M}} \circ \mathbf{Z}$, where $\circ$ denotes element-wise multiplication.

\item Time-Domain Reconstruction: The inverse Fourier transform (IFFT) is applied to $\mathbf{Y}$ to obtain a time-domain series. The desired fGn series $\mathbf{X}$ corresponds to the first $N$ points of the real part of this result, as the imaginary component should be negligible due to numerical precision errors.
\end{enumerate}

The resulting time series $\mathbf{X}$ is a stationary fractional Gaussian noise process characterized by a constant Hurst exponent H. When integrated, it exhibits the monofractal behavior of a fractional Brownian motion, making it a reliable benchmark for algorithms that analyze long-range dependence.

\subsubsection{Multifractal Series due to Fat-Tailed PDF}

The first method to generate multifractality due to value distribution involves a two-step process. First, a monofractal series with linear correlations, such as fractional Gaussian noise (fGn) described before, is generated using the Davies-Harte method. This series is characterized by a constant Hurst exponent $H$ and a Gaussian PDF. Second, a nonlinear transformation, such as raising each data point to a power, is applied. This operation preserves the underlying linear correlations but introduces a non-linear effect that significantly alters the PDF, yielding a series with a heavy-tailed distribution. The resulting multifractality is therefore a direct consequence of this broad value distribution.

\subsubsection{Multifractal Series due to Long-Range Correlations}
The canonical model for multifractality driven by non-linear correlations is the binomial multiplicative cascade. This iterative process constructs a time series with a hierarchical, multiplicative structure. For a series of length $N=2^k$, the algorithm proceeds as follows:

\begin{enumerate}
    \item An initial interval is divided into two sub-intervals, each receiving a weight (or multiplier) of either $a$ or $1-a$, where $a \in (0,1)$ and $a \neq 0.5$.
    \item This process is repeated $k = \log_2(N)$ times, with each sub-interval recursively partitioned and its values multiplied by either $a$ or $1-a$.
\end{enumerate}

While this method successfully generates a complex correlation structure, its simplest implementation has a significant side effect: the multiplication of random variables tends to produce a final PDF that is log-normal and thus heavy-tailed, rather than Gaussian. This confounds the analysis because the resulting series has both a non-linear correlation structure and a broad PDF, making it impossible to isolate the source of multifractality using a simple shuffling test.

To overcome this issue and ensure that multifractality is exclusively due to the correlation structure, we employ a method that disentangles the PDF from the correlations. This involves creating a surrogate series that maintains the precise multifractal correlations of the cascade while enforcing a perfect Gaussian distribution. The algorithm is implemented as follows:

\begin{enumerate}
    \item A multifractal cascade series is generated using the standard binomial method. This series possesses the desired long-range correlation structure.
    \item A series of perfect Gaussian white noise is generated. This series contains the desired Gaussian PDF.
    \item The rank order of the cascade series is used to sort the values of the Gaussian series. In effect, the correlation structure (the ranking) of the cascade is "imprinted" onto the Gaussian values.
\end{enumerate}

The final output is a series that retains the intricate, multifractal correlation structure of the cascade model while possessing a strictly Gaussian PDF. This allows for an unambiguous test of the hypothesis that multifractality arises from non-linear correlations, as any observed multifractal signature cannot be attributed to the value distribution. A standard test like Iterated Amplitude Adjusted Fourier Transform (IAAFT) or a simple shuffling procedure on this series will destroy its multifractal signature, confirming that its origin was indeed the long-range correlations.

\subsection{Series with a Crossover in $F_q(s)$ Functions}

We have included a computationally efficient technique that aims to generate a time series with the specific spectral properties of an fGn. The method is described in \cite{kantelhardt2001} by Kantelhardt et al. and we will refer to it as Fourier Filtering Method (FFM). The underlying concept is to mold or filter the power spectrum of a white noise signal in the frequency domain to match the theoretical power spectral density (PSD) of an fGn.

The theoretical foundation relies on the Wiener-Khinchin Theorem, which links the PSD, $S(f)$, of a stationary process to the Fourier transform of its autocorrelation function. For fGn, the PSD follows a power-law relationship:

$$ S(f)\propto \rvert f \rvert ^{-\beta}$$
 
where $f$ is the frequency and $\beta$ is the spectral exponent, directly related to the Hurst exponent ($H$) by the equation $\beta=2H-1$. Theoretically, the $\alpha$ exponent measured by DFA in an fGn series is equivalent to its Hurst exponent $H$.

The generation algorithm is rapid, typically operating with a computational complexity of $O(N\,logN)$ due to the reliance on the Fast Fourier Transform (FFT). However, FFM is recognized as an approximate method. The intrinsic discretizations of the FFT, alongside implementation choices, causes the generated series' spectrum to deviate from the ideal theoretical spectrum, particularly at the extreme boundaries of the frequency domain (the lowest and highest frequencies). This practical limitation often manifests as the aforementioned crossover behavior in the $F_q(s)$ functions, making the FFM an ideal benchmark for testing the sub-range fitting capabilities of our MFDFA implementation. A notable limitation arises when generating highly anti-persistent series (low Hurst exponents, $H<0.5$) due to the method's reliance on the Power Spectral Density (PSD), $S(f)$, which requires the spectrum to grow with frequency, an implementation constraint that can compromise the fidelity of the generated signal.

\vspace{2mm}

The Fourier Filtering Method (FFM) is implemented through the following steps for a desired series length $N$:

\begin{enumerate}
    \item Generate White Noise: A time series $x(t)$ of $N$ points is created from a standard Gaussian distribution (zero mean, unit variance).
    \item Fourier Transform: The Fast Fourier Transform (FFT) of the white noise signal is computed: $X(f) = \mathcal{F}\{x(t)\}$.
    \item Construct the Filter: A filter $F(f)$ is defined in the frequency domain based on the theoretical Power Spectral Density (PSD), $S(f)$, of a fractional Gaussian noise. Since power is proportional to the amplitude squared, the amplitude filter is the square root of the PSD:
    \begin{equation}
        F(f) \propto \sqrt{S(f)} \propto |f|^{-\beta/2} = |f|^{-(2H-1)/2}
    \end{equation}
    \item Apply the Filter: The spectrum of the white noise is multiplied by the constructed filter, effectively shaping the noise's frequency content:
    \begin{equation}
        Y(f) = X(f) \cdot F(f)
    \end{equation}
    \item Inverse Transform: The Inverse Fast Fourier Transform (IFFT) is calculated to obtain the final time-domain series $y(t)$ with the desired spectral properties:
    \begin{equation}
        y(t) = \mathcal{F}^{-1}\{Y(f)\}
    \end{equation}
\end{enumerate}

\subsection{Methodological Guidelines and Default Parameters}

While the \texttt{MF-toolkit} library optimizes and automates the rigorous computation of the MFDFA and crossover detection algorithms, it is crucial to emphasize that software automation does not eliminate all methodological choices. The accuracy of multifractal analysis remains inherently sensitive to the user's parameterization of the fluctuation function. To ensure robust analysis, we recommend the following heuristic guidelines and parameter constraints:

\begin{itemize}
    \item \textbf{Moment exponent range ($q$):} We advise avoiding extreme $q$ values when analyzing short time series. While high positive or negative $q$ exponents are necessary to amplify macroscopic and microscopic fluctuations respectively, they are highly susceptible to divergence and statistical instability driven by finite-size effects. A standard default range of $q \in [-5, 5]$ or $q \in [-10, 10]$ is recommended, provided the dataset contains at least $N \ge 10^4$ points.
    
    \item \textbf{Scale range ($s$):} The selection of the scaling window must strictly adhere to theoretical bounds. To avoid discretization errors and artificial correlations at microscopic scales, the minimum scale should generally be bounded by $s_{min} \ge 10$ (or strictly $s_{min} > m+2$, where $m$ is the detrending order). Conversely, to ensure sufficient statistical averaging across the non-overlapping segments, the maximum scale must not exceed $s_{max} \le N/4$, with $s_{max} \le N/5$ being a safer default for highly non-stationary data.
    
    \item \textbf{Polynomial order ($m$):} The detrending order $m$ dictates the complexity of the local polynomial removed from each segment. Guidelines suggest selecting $m=1$ (linear) or $m=2$ (quadratic) based on the inherent macroscopic non-stationarity of the signal. While higher orders ($m \ge 3$) can successfully eliminate more complex low-frequency trends, they introduce a severe risk of overfitting at small scales, which artificially suppresses the true fluctuation function $F_q(s)$ and skews the Hurst exponent estimation.
\end{itemize}
\section{Results and Discussion}

\label{sec:results}
The \texttt{MF-toolkit} library is built upon standard scientific Python packages (\texttt{NumPy}, \texttt{SciPy}) and leverages \texttt{Numba} for parallelization and just-in-time compilation to accelerate critical calculations.
\subsection{Validation of Algorithmic Innovations using Synthetic Data}
\subsubsection{Generating synthetic time series}

For this analysis, we have generated three representative time series using MF-Toolkit: i) a monofractal series, typically a fractional Brownian motion; ii) a multifractal series whose source is a broad probability distribution function with fat-weighted tails; and iii) a multifractal series characterized by its long-range correlations. 

\begin{figure}[h!]
    \centering
    \includegraphics[width=\linewidth]{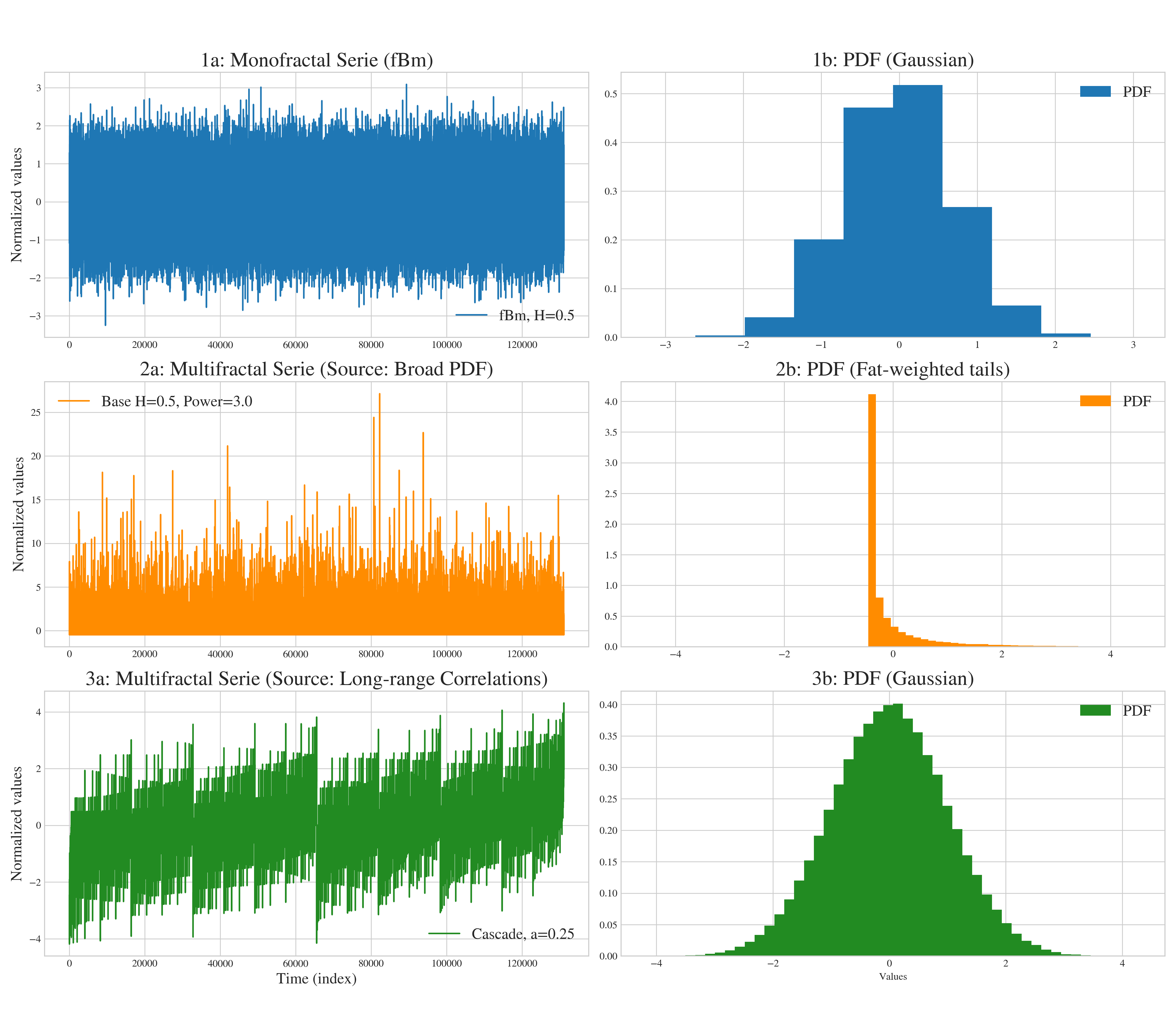}
    \caption{\textbf{Representative Synthetic Time Series and Corresponding Probability Distribution Functions}. Panels (1a) and (1b) show a monofractal Fractional Brownian Motion (fBm) series (H=0.5) and its Gaussian PDF, respectively. Panel (2a) illustrates a series with multifractality originating from a heavy-tailed distribution, shown in its PDF (2b); and (3a) depicts a series where multifractality stems from long-range correlations, with its corresponding Gaussian PDF in (3b).}
    \label{fig:synthetic_series}
\end{figure}

In Figure \ref{fig:synthetic_series} could be seen the three cases with its respective histograms, which depict the shapes of their probability distribution functions (PDF). 

\begin{figure}[h!]
    \centering
    \includegraphics[width=0.9\linewidth]{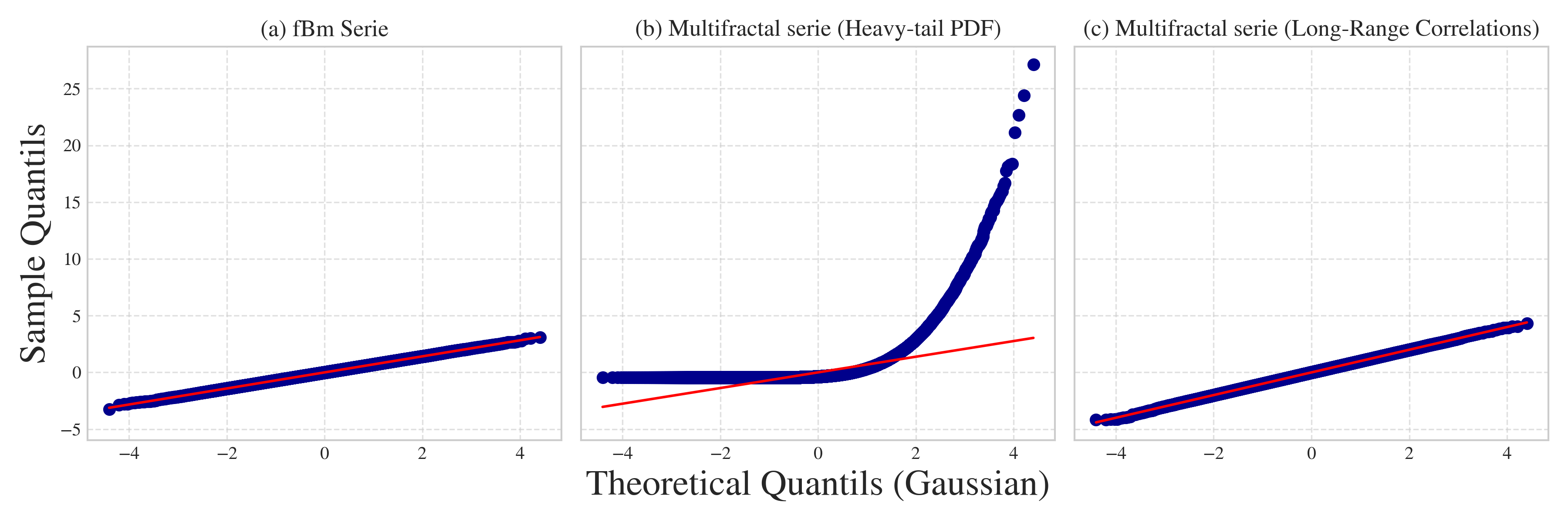}
    \caption{\textbf{Quantile-Quantile (Q-Q) Plots of Synthetic Time Series against a Standard Gaussian Distribution}. (a) fBm Series: The data points closely follow the Gaussian reference line, confirming that the monofractal Fractional Brownian Motion (fBm) series exhibits a Gaussian distribution of values. (b) Multifractal Series (Heavy-tail PDF): A significant deviation from the linear reference is observed, particularly in the extreme quantiles (tails). (c) Multifractal Series (Long-Range Correlations): The majority of data points align closely with the Gaussian reference line.}
    \label{fig:qqplots}
\end{figure}

For comparing the distribution of each case to a standard Gaussian distribution we proposed to assess a Q-Q plot. This is a typical visual inspection in statistics where the theoretical quantiles of a normal distribution are compared with the quantiles of the sample under analysis. If the distribution were a standard Gaussian it must follows an identity function.
Figure \ref{fig:qqplots} shows the Q-Q plots for each series and complements Figure \ref{fig:synthetic_series} by providing a more detailed statistical visualization of the underlying value distributions for each synthetic series. All Q-Q plots compare the empirical quantiles of the respective time series against the theoretical quantiles of a standard Gaussian distribution (red line represents perfect agreement).

The plots show an evident deviation in the case \ref{fig:qqplots}(b) obtaining a typical plot for a heavy-tailed distribution, with more probability of abnormal events occurring. The pronounced S-shape confirms the presence of heavy tails in the value distribution, which is the source of multifractality for that series.

\subsubsection{Multifractal Detrended Fluctuation Analysis of Synthetic Time Series}

In this section we apply the MFDFA function developed in MF-Toolkit library to assess multifractal signatures of multifractal synthetic signals.

The developed Multifractal Detrended Fluctuation Analysis (MFDFA) function incorporates several key features to enhance computational performance and methodological precision. Firstly, it leverages parallel processing to significantly accelerate the fluctuation function calculation. By distributing the independent computations for each moment $q$ across multiple CPU cores, the library drastically reduces execution time, making it highly efficient for analyzing large datasets and for large-scale studies requiring extensive parameter sweeps. Users can set the number of cores to utilize with a simple parameter.

\vspace{2mm}

Secondly, the library offers an optional sub-range fitting capability for determining the generalized Hurst exponent $h(q)$. This feature allows users to exclude anomalous behaviors that often occur at very small or very large scales. This is particularly important for accurately characterizing the scaling behavior, as MFDFA is a method that requires a well-defined scaling region. By allowing the user to precisely select the linear fitting range, the library ensures a more robust estimation of the multifractal exponents, improving the reliability and clarity of the results.

\vspace{2mm}

Finally, the function includes a built-in validation filter that can be toggled on or off with a simple validate boolean flag. This feature performs a series of theoretical and numerical checks on the input data and the computed results. For example, it can verify if the singularity spectrum, $f(\alpha)$, is in range [0,1] and if its concavity is negative. This helps prevent common pitfalls and provides a layer of quality control, guiding users toward sound methodological practices and ensuring the validity of the final multifractal analysis.

\vspace{2mm}

\begin{figure}[h!]
    \centering
    \includegraphics[width=\linewidth]{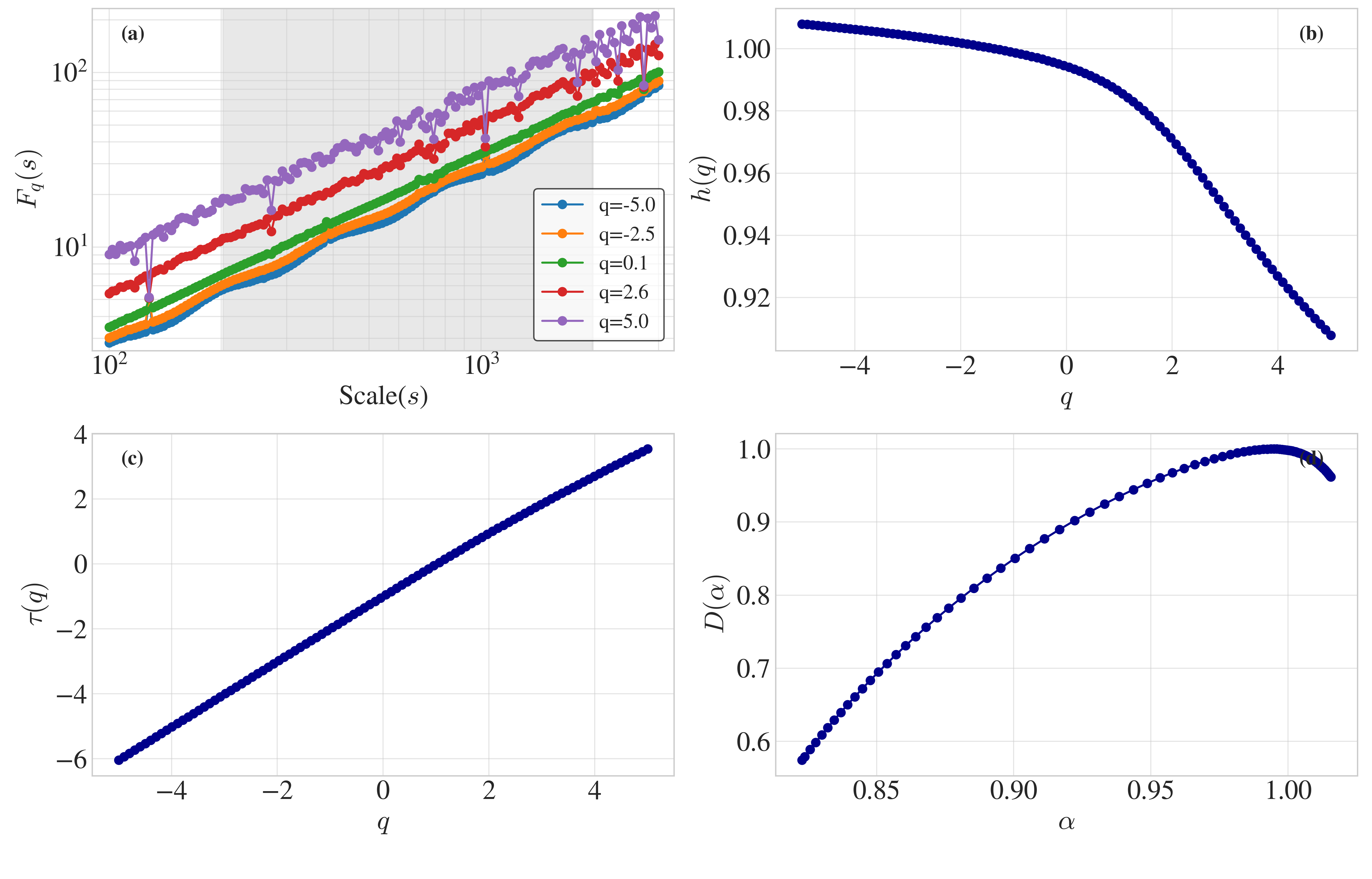}
\caption{\textbf{MFDFA Curves for Synthetic Multifractal Signal (Long-Range Correlations)}. The complete Multifractal Detrended Fluctuation Analysis workflow. (a) Generalized fluctuation function $F_q(s)$ vs. scale $s$; (b) Generalized Hurst exponent $h(q)$ vs. moment $q$; (c) Mass exponent $\tau(q)$ vs. $q$; (d) Singularity spectrum $f(\alpha)$ vs. $\alpha$. The non-constant $h(q)$ confirms multifractal scaling.}

\label{fig:mfdfa_corr}
\end{figure}

\begin{figure}[h!]
    \centering
    \includegraphics[width=\linewidth]{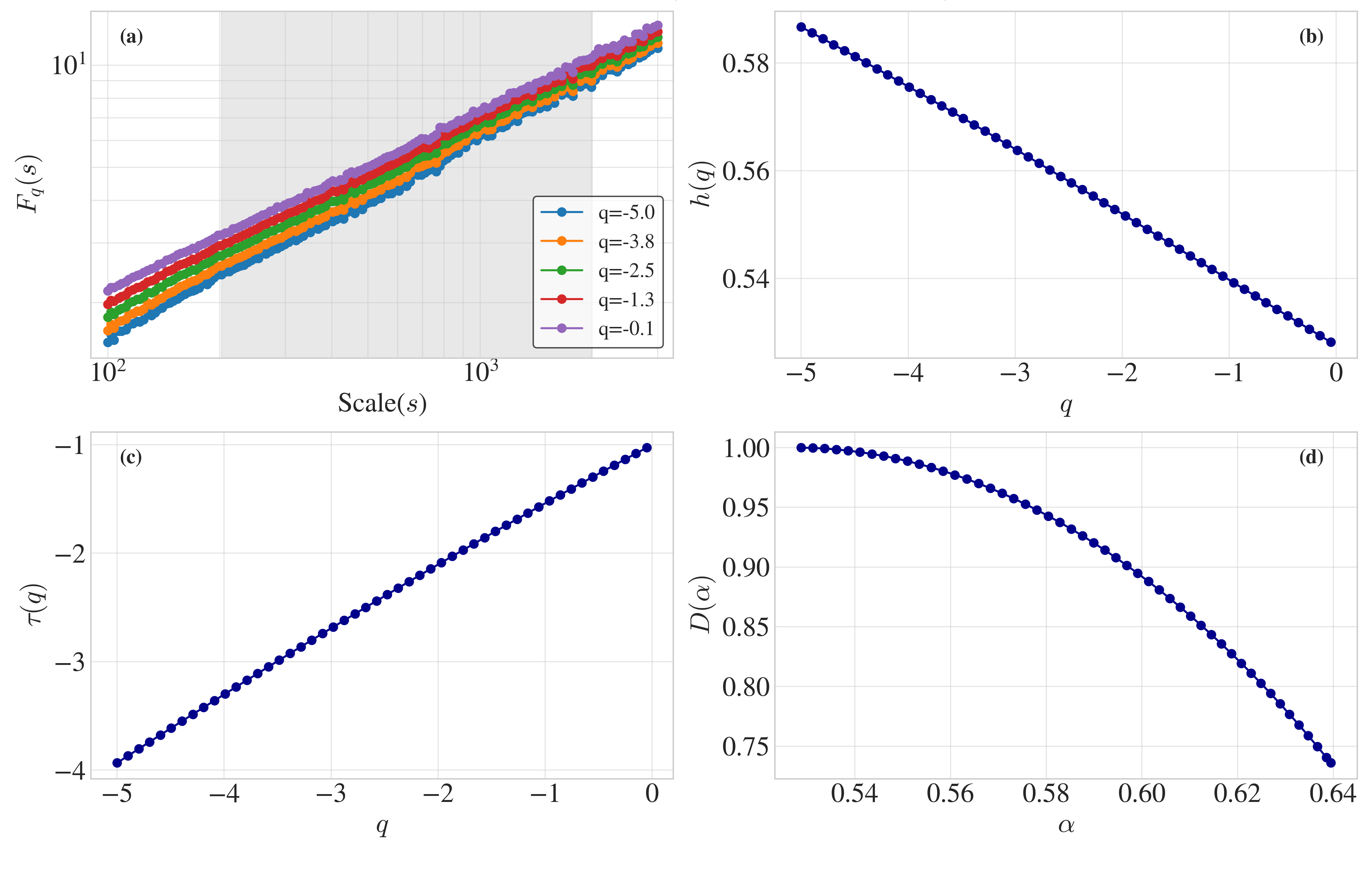}
   \caption{\textbf{MFDFA Curves for Synthetic Multifractal Signal (Heavy-Tailed Distribution)}. The MFDFA analysis. (a) Generalized fluctuation function $F_q(s)$; (b) Generalized Hurst exponent $h(q)$; (c) Mass exponent $\tau(q)$; (d) Singularity spectrum $f(\alpha)$. Note the dependence of $h(q)$ on $q$, which is characteristic of multifractality originating from the heavy-tailed value distribution.}
\label{fig:mfdfa_dist}
\end{figure}

Figures \ref{fig:mfdfa_corr} and \ref{fig:mfdfa_dist} illustrate the characteristic plots for multifractal series generated from distinct sources. The four panels illustrate the full workflow of a Multifractal Detrended Fluctuation Analysis (MF-DFA). The results confirm that both synthetic series exhibit a clear multifractal signature, which is consistent with their method of generation.

\newpage
\subsubsection{Source Identification with Surrogate Data}
According to \cite{Kantelhardt2002} the multifractality in a time series can arise from two distinct sources: i) Long-range correlations: Dependencies between values of the series at different times. If these correlations are nonlinear (e.g., volatility clustering, where the magnitude of fluctuations is correlated), they can induce different scaling for different statistical moments $q$, leading to multifractality.
\\
ii) A broad probability distribution function of the series' values. If the PDF is heavy-tailed (e.g., Lévy or Pareto distributions), large fluctuations are much more probable than in a Gaussian process. These rare, extreme events can dominate the sum in $F_q(s)$ for large positive $q$, while being less important for small or negative $q$, thus creating a $q$-dependent $h(q)$.

\vspace*{2mm}

To determine the origin of multifractality, \texttt{MF-toolkit} includes methods to generate surrogate data that preserve certain properties of the original series while destroying others.

\begin{itemize}
    \item \textbf{Shuffling:} Randomly permutes the time series values. This destroys all temporal correlations (both linear and nonlinear) but perfectly preserves the PDF.
    \item \textbf{IAAFT (Iterative Amplitude Adjusted Fourier Transform):} A more sophisticated method that generates a surrogate series preserving both the PDF (amplitude distribution) and the power spectrum (linear correlations) of the original series. It does so by iteratively adjusting the amplitudes in the Fourier domain and the values in the time domain. This method effectively destroys any nonlinear correlations.
\end{itemize}

The logic of the test is as follows: If the multifractality persists in the shuffled surrogates, its origin is the heavy-tailed PDF. If the multifractality vanishes in the shuffled surrogates but the original series is multifractal, the origin is in the correlations. Moreover if the multifractality vanishes in the IAAFT surrogates, its origin is specifically in the \textit{nonlinear} correlations.

We have generated surrogate time series applying the random shuffle technique discussed before 50 times and then taking the mean values. 

\begin{figure}[h!]
    \centering
    \includegraphics[width=0.9\linewidth]{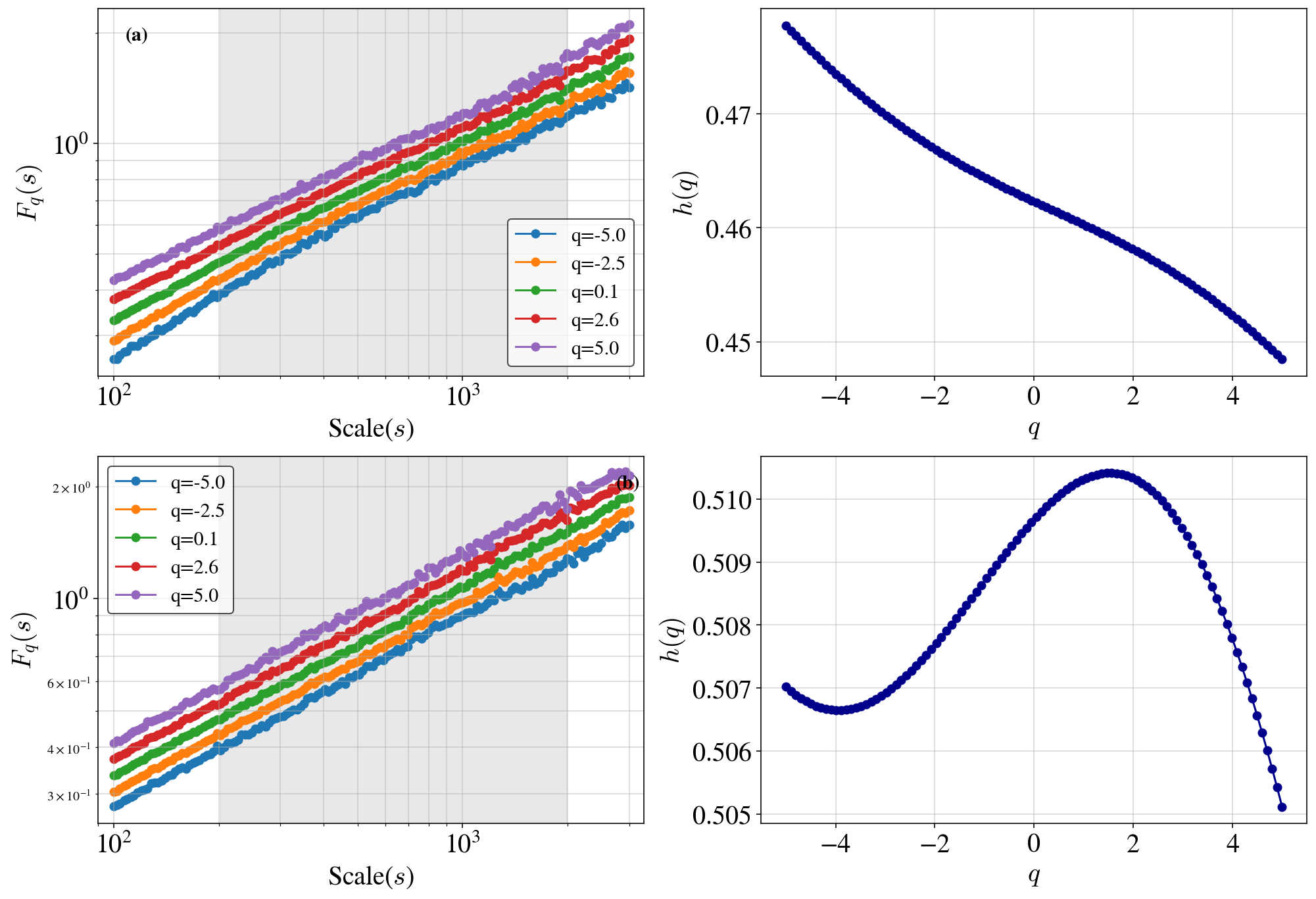}
    \caption{\textbf{The Effect of Random Shuffling on Multifractal Signatures}. Panels (a) and (b) show the generalized fluctuation functions, $F_q (s)$, and the generalized Hurst exponent, $h(q)$, respectively, for the multifractal series where multifractality is due to a heavy-tailed probability distribution. Panels (c) and (d) show the corresponding plots for the multifractal series where multifractality is due to long-range correlations.}
    \label{fig:shuffled_series_mfdfa}
\end{figure}

Applying MFDFA to a shuffled surrogate of the broad-PDF series reveals that its multifractality remains intact. This is because the shuffling procedure preserves the series' heavy-tailed value distribution, which is the sole source of its multifractality. In contrast, the multifractality of the correlation-based series is completely destroyed by shuffling. The resulting generalized Hurst exponent, $h(q)$, is a constant $0.5$ for all values of $q$, which is the signature of an uncorrelated, monofractal signal. This confirms that the multifractality of this series was exclusively due to its long-range correlation structure. This is illustrated in Figure \ref{fig:shuffled_series_mfdfa}.

To validate the source of multifractality in each synthetic series, we generated 10 surrogate time series for each signal using the Iterated Amplitude Adjusted Fourier Transform (IAAFT) technique and then averaged the results. The IAAFT method effectively destroys nonlinear correlations while preserving both the linear correlation structure and the probability distribution of the original signal.

\vspace{2mm}

After applying the MFDFA, we observed a crucial difference between the two series. For the signal where multifractality was due to long-range correlations, the shuffling procedure completely eliminated the multifractal signature. This indicates that its multifractality was driven by a nonlinear, rather than linear, correlation structure. In contrast, the multifractal signature of the series with a heavy-tailed PDF remained almost entirely intact, as the shuffling preserved its value distribution, which was the sole source of its multifractality.

\begin{figure}[h!]
    \centering
    \includegraphics[width=0.9\linewidth]{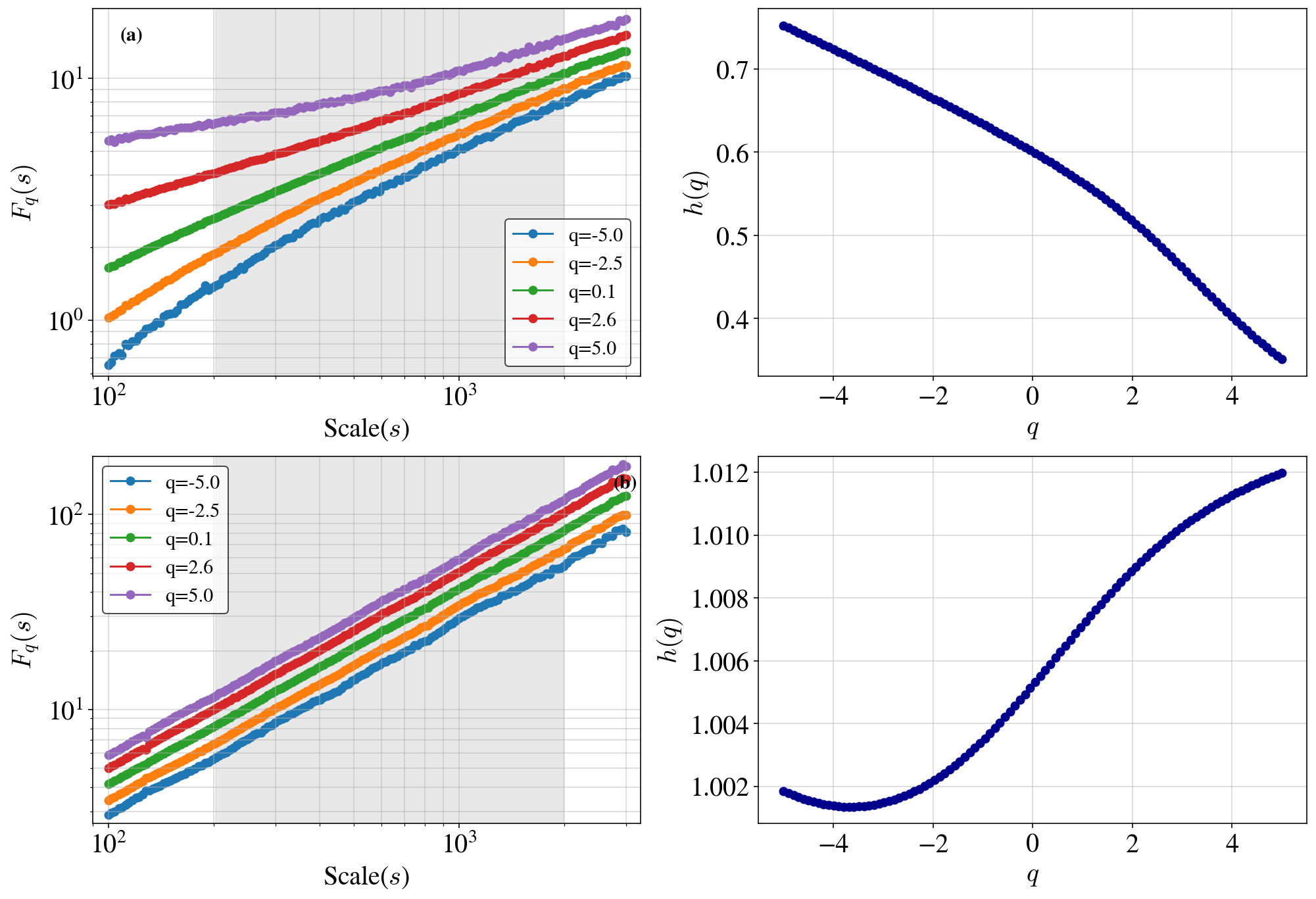}
    \caption{\textbf{The Effect of IAAFT Shuffling on Multifractal Signatures}. Panels (a) and (b) show the generalized fluctuation functions, $F_q (s)$, and the generalized Hurst exponent, $h(q)$, respectively, for the multifractal series where multifractality is due to a heavy-tailed probability distribution. Panels (c) and (d) show the corresponding plots for the multifractal series where multifractality is due to long-range correlations.}
    \label{fig:iaaft_series_mfdfa}
\end{figure}

\newpage

\subsubsection{Generating Fluctuation Functions with crossovers}
To effectively validate our algorithm's capability in identifying scaling transitions, we generated a synthetic time series that deliberately exhibits a crossover at a specified scale index. This series was created using the Fourier Filtering Method (FFM), as described in Section \ref{sec:theory}. We implemented a function within the FFM framework that allows for the imposition of two distinct scaling regimes, characterized by separate Hurst exponents ($H_1$ for the short-range behavior and $H_2$ for the long-range behavior). This controlled generation process provides an ideal benchmark for testing the precision of our library's sub-range fitting feature in isolating and quantifying different scaling exponents.

\begin{figure}[h!]
    \centering
    \includegraphics[width=\linewidth]{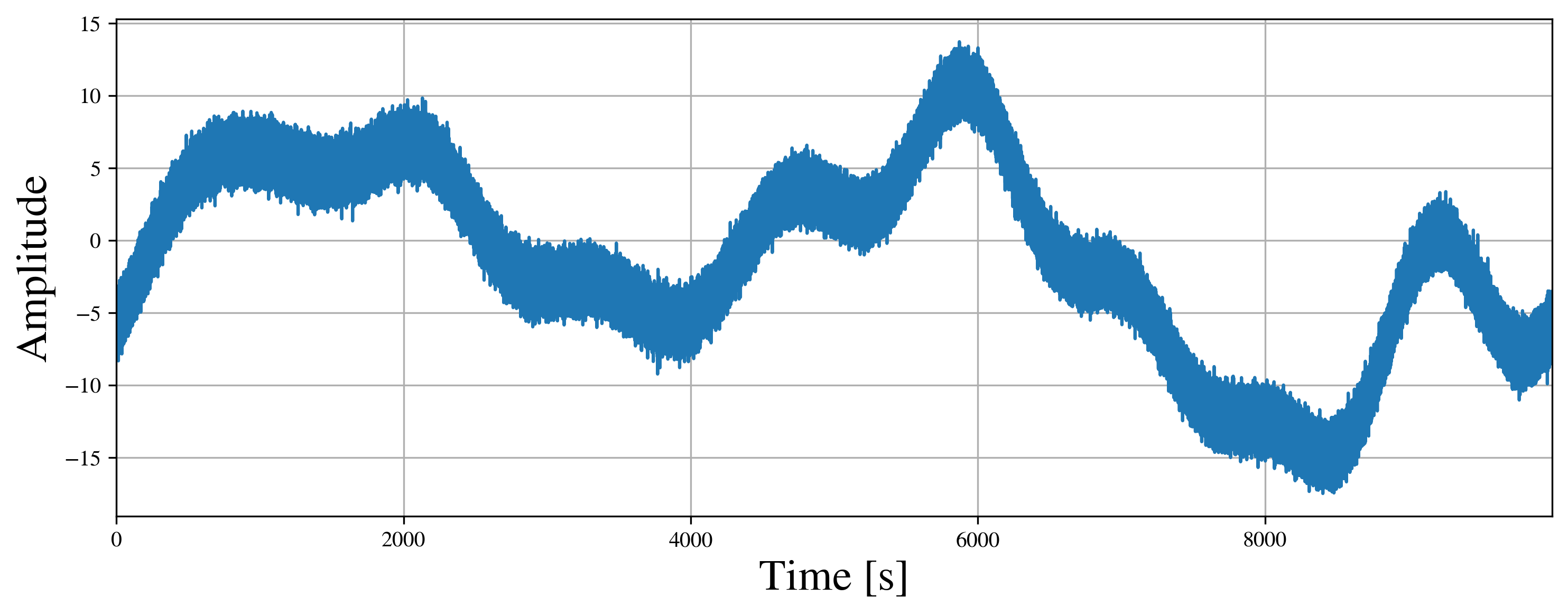}
    \caption{\textbf{Synthetic Time Series Generated using the Fourier Filtering Method (FFM)}. This series was generated using the FFM algorithm to exhibit a crossover behavior, resulting from the imposition of two distinct scaling regimes. The generated fractional Gaussian noise (fGn) process is characterized by two designated Hurst exponents: $H_1$ for the long-range behavior and $H_2$ for the short-range behavior.}
    \label{fig:generate_serie_crossover}
\end{figure}

\newpage

\subsubsection{Automated Crossover Detection}
\begin{figure}[h!]
    \centering
    \includegraphics[width=\linewidth]{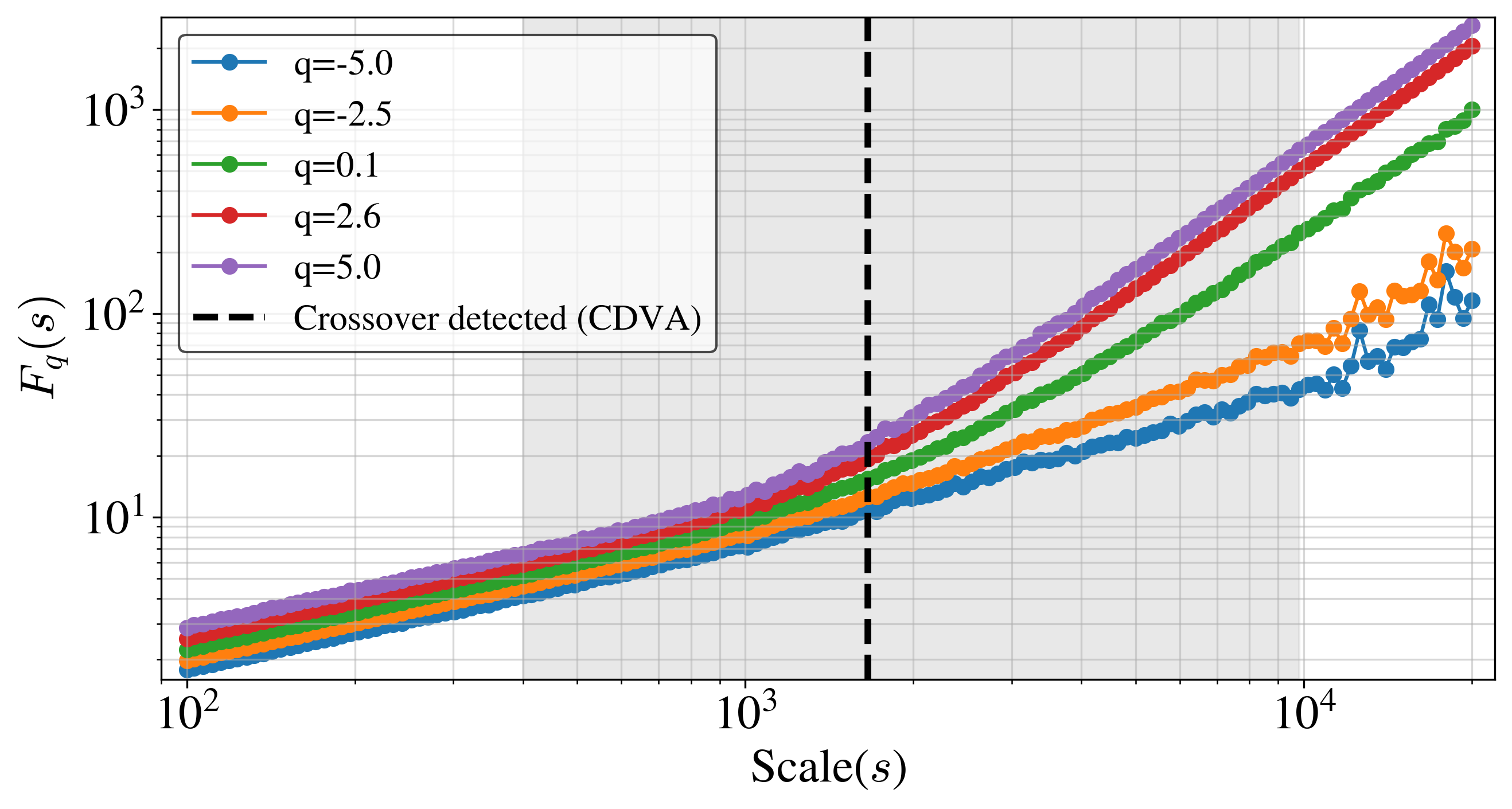}
    \caption{\textbf{Crossover Detection in $\mathbf{F_q(s)}$ Functions using the CDV-A Method}. This figure illustrates the application of the Crossover Detection and Validation Algorithm (CDV-A) to the generalized fluctuation functions, $F_q(s)$, of the synthetic series with two scaling regimes. The dashed vertical line indicates the crossover index detected automatically by the CDV-A, determined by averaging the detection results across all statistical moments $q$.}
    \label{fig:cdva_crossover}
\end{figure}

To automate the challenging task of identifying scaling transitions, we implemented the Crossover Detection based on Variance of slopes differences Algorithm (CDV-A) proposed by Moreno-Pulido et al. Following the application of the MFDFA method to the synthetic time series generated in the previous section, the CDV-A automatically identifies the crossover index. For a robust estimation of the crossover point, users have the option to either average the detection index across all statistical moments q or to rely solely on the $q=2$ case (corresponding to Detrended Fluctuation Analysis, DFA), which offers a simplified yet often sufficient estimate. In Figure \ref{fig:cdva_crossover} results of the detection are shown for $q=2$ case.

In addition, we have incorporated the Sequential Permutation fo Identifying Crossovers (SPIC) algorithm, detailed in Section \ref{sec:theory}. This robust method offers a methodological advantage over conventional techniques, such as the widely used CDVA method, by enabling the detection of multiple crossover points. However, it is important to note that this increased precision necessitates a higher computational cost, which must be managed by the user.

\begin{figure}[h!]
    \centering
    \includegraphics[width=\linewidth]{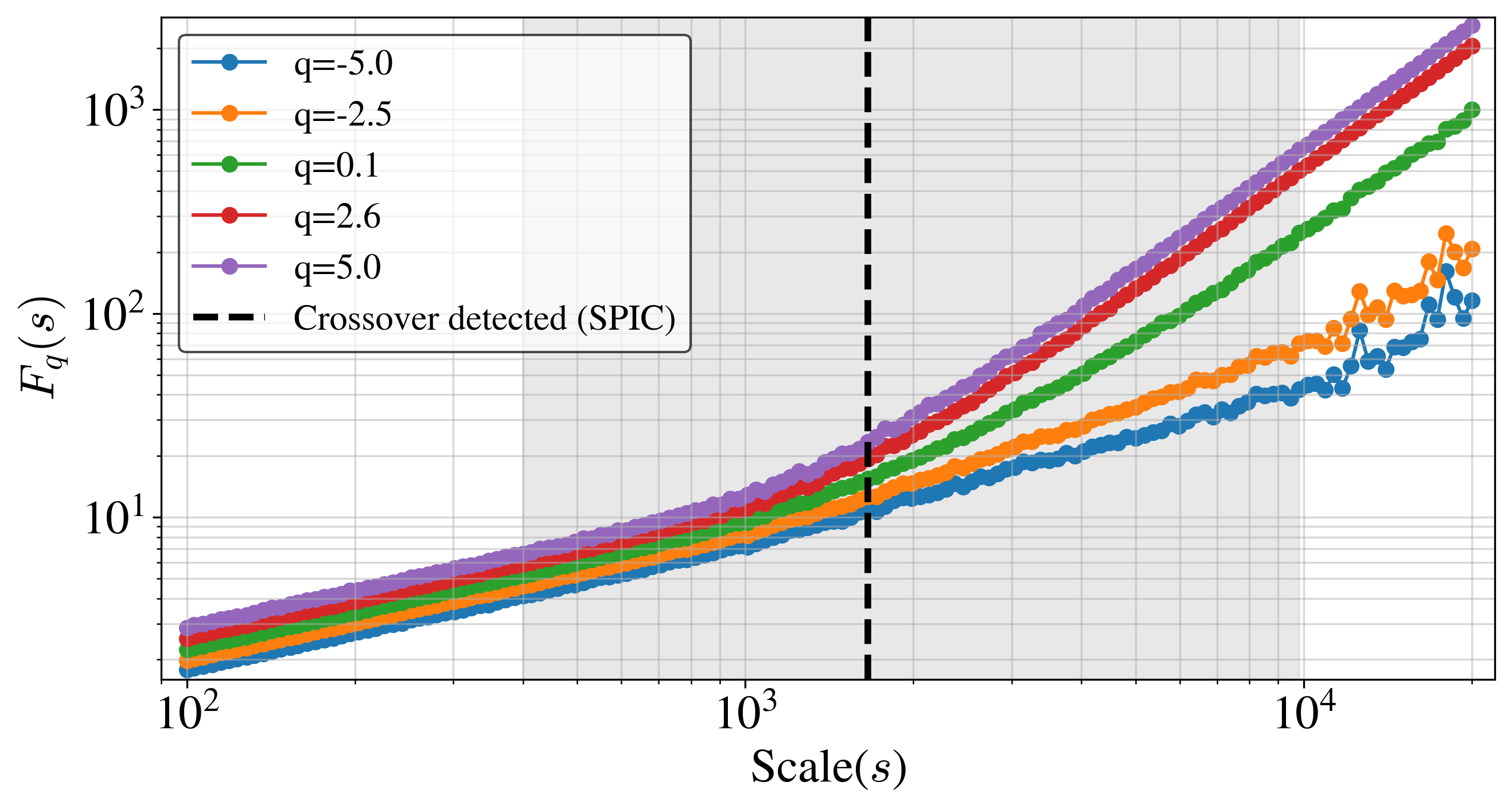}
    \caption{\textbf{Crossover Detection in $\mathbf{F_q(s)}$ Functions using the SPIC Method}. This figure illustrates the application of the Sequential Permutation for Identifying Crossovers (SPIC) algorithm to the generalized fluctuation functions, $F_q(s)$, of the synthetic series with two scaling regimes. The dashed vertical line indicates the crossover index detected automatically by the SPIC algorithm.}
    \label{figg:spic_crossover}
\end{figure}

\begin{figure}[h!]
    \centering
    \includegraphics[width=\linewidth]{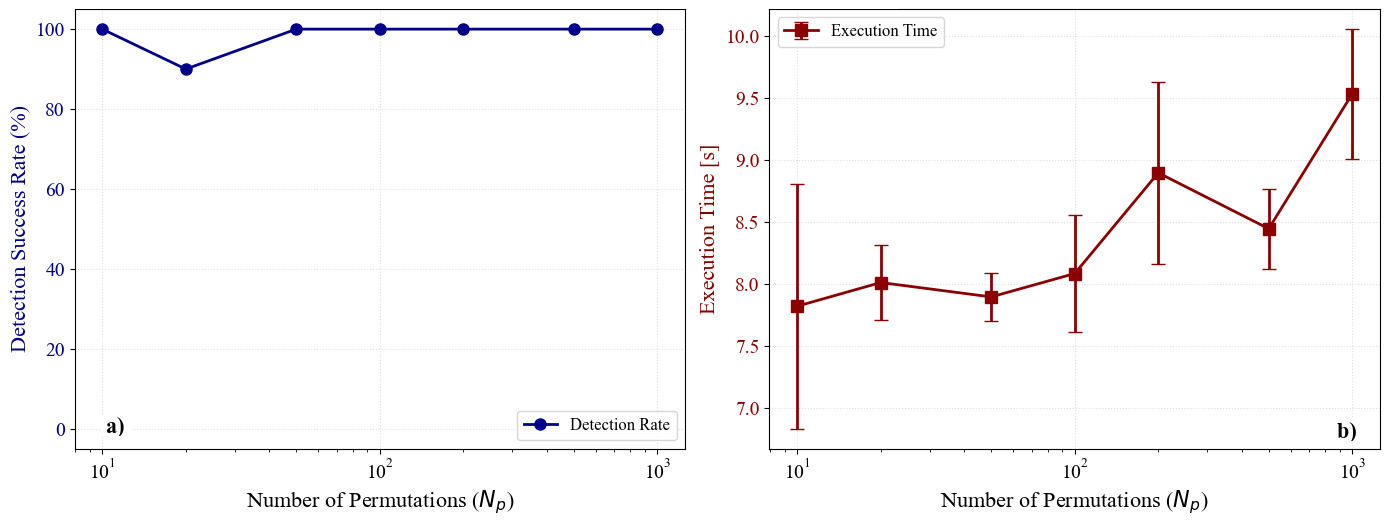}
    \caption{\textbf{SPIC performance and statistical convergence under non-ideal conditions.} The algorithm was tested on a synthetic series ($N=100000$) with a subtle structural break ($\alpha_1=0.9$, $\alpha_2=1.1$) and 30\% additive Gaussian noise. \textbf{a)} Statistical reliability: Rate of true detections over 10 independent trials. \textbf{b)} Computational cost: Mean execution time, where error bars denote $\pm 1$ standard deviation over the 10 independent trials.}
\label{fig:spic_benchmark}
\end{figure}

To empirically determine the optimal number of permutations ($N_p$) for the SPIC algorithm, we evaluated its detection reliability and computational cost under non-ideal conditions: a synthetic  time series with subtle crossover with 30\% additive noise. As shown in Fig. \ref{fig:spic_benchmark}a, low values of $N_p$ (e.g., $N_p=20$) can lead to false negatives (90\% success rate) due to insufficient statistical resolution when evaluating the $p$-value. However, convergence is rapidly achieved, with the detection rate stabilizing at 100\% for $N_p \ge 100$. Remarkably, as depicted in Fig. \ref{fig:spic_benchmark}b, the execution time exhibits a very gentle slope. This minimal marginal cost for additional permutations is a direct consequence of the low-level C-vectorization and multiprocessing architecture. The fixed overhead of thread initialization dominates the execution, while the mathematical permutations themselves are evaluated almost instantaneously. Consequently, setting $N_p$ between 100 and 200 provides an optimal operational balance, guaranteeing absolute statistical robustness without a prohibitive computational penalty.

\subsubsection{Quantitative Robustness to Noise}
To evaluate the robustness and statistical stability of the crossover detection methods under non-ideal experimental conditions, we conducted a Monte Carlo stress test. We systematically corrupted a synthetic baseline signal by superimposing additive Gaussian white noise. The noise intensity was progressively scaled so that its standard deviation corresponded to 0\%, 5\%, 10\%, 20\% and 30\% of the original signal's standard deviation. For each discrete noise level, an ensemble of independent stochastic realizations was generated and subsequently analyzed using both the CDVA and SPIC algorithms to track the accuracy and variance of the detected crossover locations.\\

\begin{figure}[h!]
    \centering
    \includegraphics[width=\linewidth]{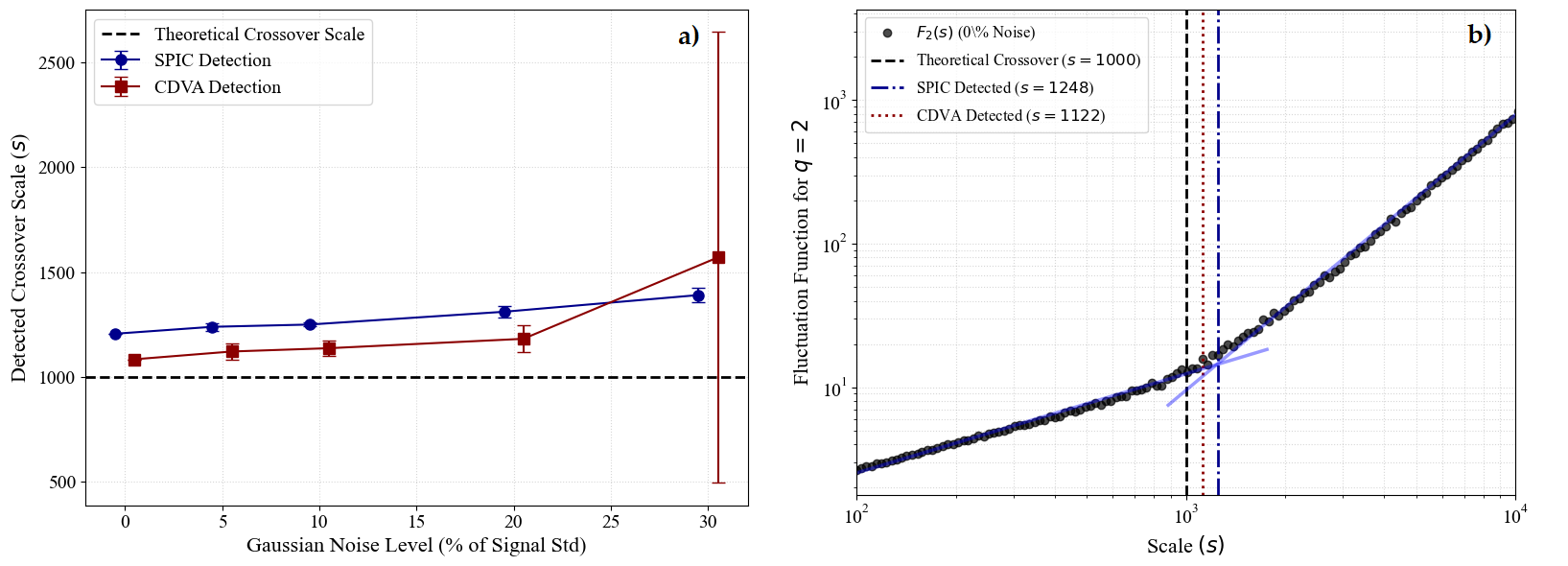}
    \caption{\textbf{Robustness of the crossover detection algorithms and systematic offset evaluation}. \textbf{a)} Monte Carlo simulations for the SPIC (blue circles) and CDV-A (red squares) algorithms applied to synthetic multifractal time series corrupted with progressive levels of additive Gaussian white noise. Markers indicate the mean detected crossover scale across 100 independent stochastic realizations for each noise level, while error bars denote $\pm 1$ standard deviation. \textbf{b)} Log-log plot of the fluctuation function $F_2(s)$ for a single clean synthetic realization (0\% noise).}
\label{fig:noise_crossover}
\end{figure}

The results of this perturbation analysis (Fig. \ref{fig:noise_crossover}a) demonstrate a clear divergence in the robustness of the two algorithms under heavy noise. While both methods successfully attenuate high-frequency white noise at macroscopic scales and maintain mean crossover values close to the theoretical expectation at low-to-moderate noise intensities (up to 20\%), their statistical stability differs drastically under severe corruption. Specifically, the CDV-A algorithm exhibits a dramatic increase in variance when the noise level reaches 30\%. This instability occurs because CDV-A relies on the local variance of slope differences, a metric that becomes highly sensitive to aggressive stochastic fluctuations. 

\vspace{3mm}

In stark contrast, the SPIC algorithm maintains a remarkably tight variance across the entire tested noise spectrum. The narrow error bars observed for SPIC, even at 30\% noise, highlight the exceptional robustness of its iterative, permutation-based sequential testing approach. By relying on robust statistical resampling rather than local geometric variances, SPIC effectively isolates true structural crossovers from background noise. However, this statistical resilience introduces a practical trade-off, as the permutation test renders SPIC significantly more computationally expensive than the variance-based CDV-A approach (see Section \ref{sec:performance}). Consequently, while CDV-A provides a rapid estimation for relatively clean data, SPIC stands as the definitive, highly reliable tool for analyzing noisy empirical datasets without the strict need for prior signal filtering.

\vspace{3mm}

Based on the combined results of statistical robustness and computational cost (Table \ref{tab:crossover_times}), we establish a clear practical guideline for users. The CDV-A method is highly recommended for the rapid, exploratory screening of large datasets, or for obtaining initial baseline estimates in relatively clean signals. In contrast, the SPIC algorithm should be the definitive choice when analyzing highly noisy empirical time series, when multiple sequential crossovers are suspected, or when rigorous statistical confidence is mandatory for hypothesis testing.

\vspace{3mm}

Finally, it is worth noting a slight systematic overestimation in the detected crossover scales across both methods compared to the theoretical baseline ($s = 1000$). As visually explained in Fig. \ref{fig:noise_crossover}b, this constant positive bias is not an algorithmic flaw, but an intrinsic artifact of the methodology. Because the structural break produces a smooth transitional curve in the fluctuation function $F_2(s)$ rather than a sharp geometric vertex, the mathematical intersection of the piecewise linear regression models naturally shifts to the right. This geometric blending effect fully accounts for the offset observed in panel (a), confirming that the algorithms capture the phenomenological break within expected mathematical constraints.

\subsection{Application to Real-World Data: Gravitational Wave Noise Characterization}
\label{sec:LIGO}

To demonstrate the capability of \texttt{MF-toolkit} to handle high-frequency, non-stationary real-world data and validate its utility in noise characterization, we performed a systematic analysis of LIGO strain data.

\subsubsection{Data Description}
We analyzed datasets from the Gravitational Waves Open Science Center \cite{GWOSC}, including GW190408 and GW190412. Each dataset consists of time series recorded by LIGO detectors (H1 and L1) at 16384 Hz. We processed 32-second intervals centered on the merger ("Event") and 32-second intervals preceding the event ("Pre-event"/Noise).

\subsubsection{Multifractal Signatures: Signal vs. Noise}
\begin{figure}[ht]
    \centering
    \includegraphics[width=1\textwidth]{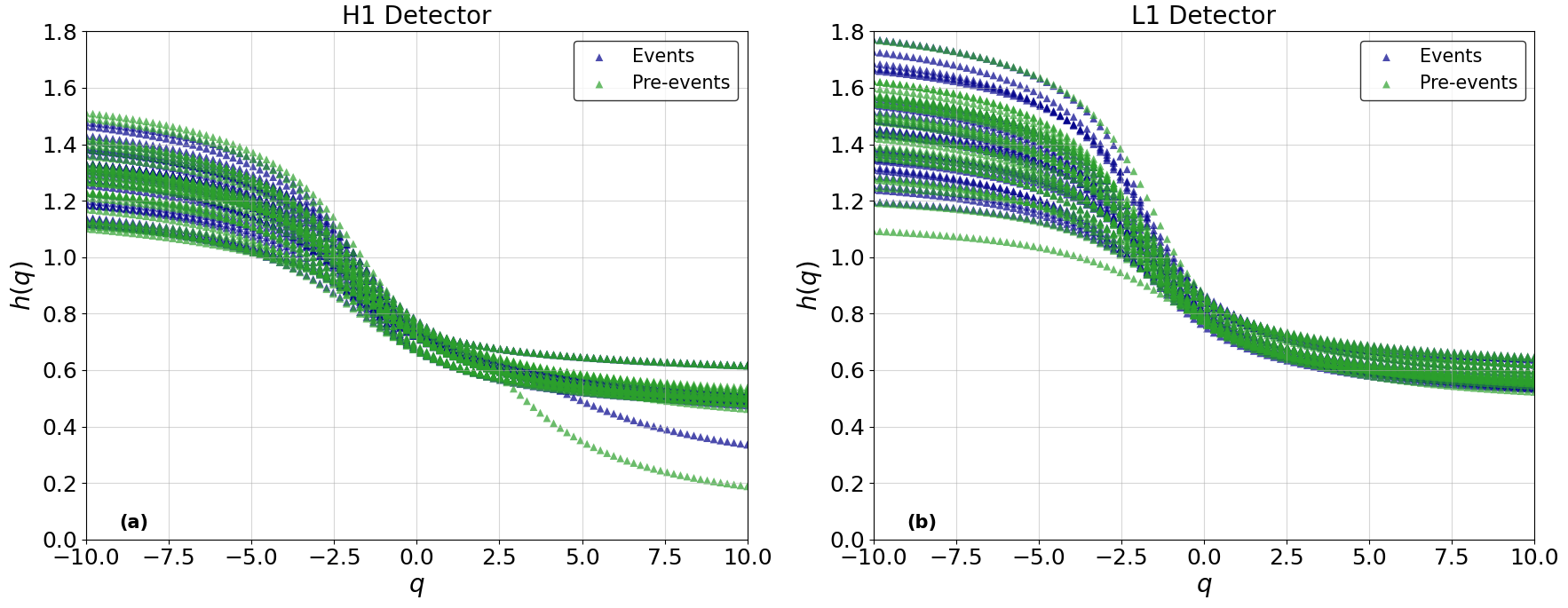}
    \caption{Generalized Hurst exponent \( h(q) \) for H1 and L1 detectors. Note the similarity between 'Events' and 'Pre-events'.}
    \label{fig:hq}
\end{figure}

We applied \texttt{mfdfa} to both event and pre-event series. Figure \ref{fig:hq} shows the generalized Hurst exponent $h(q)$. While a monotonic decrease indicates multifractality, the curves for the events overlap significantly with those of the background noise.

\begin{figure}[h!]
    \centering
    \includegraphics[width=1\textwidth]{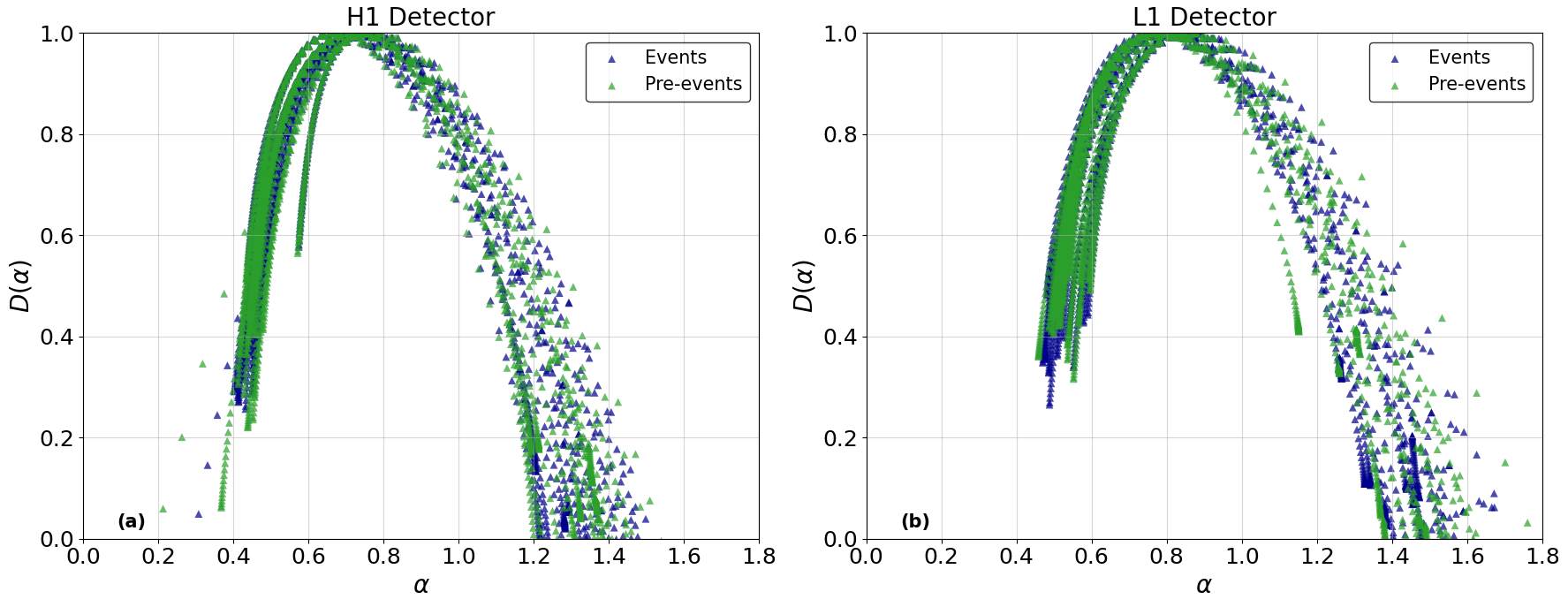}
    \caption{The multifractal spectrum $D(\alpha)$. The overlapping curves indicate that events do not possess a distinct multifractal signature compared to noise.}
    \label{fig:D}
\end{figure}

This result suggests that the observed multifractality is dominated by the detector noise, obscuring the astrophysical signal.

\subsubsection{Detector-Specific Discrepancies}
Comparing the two detectors, L1 consistently exhibits a broader multifractal spectrum than H1 (Fig. \ref{fig:D}). Table \ref{tab:statistics_mf} summarizes the parameters, showing negligible difference between "event" and "noise" phases but significant difference between detectors.

\begin{table}[h!]
\centering
\caption{Statistical description of multifractal features showing detector dependence.}
\label{tab:statistics_mf}
\begin{tabular}{|c|ll|ll|ll|ll|}
\hline
  \textbf{Category}& \multicolumn{4}{|c|}{\textbf{Events}} & \multicolumn{4}{c|}{\textbf{Pre-events}} \\ \hline
\textbf{Detector} & \multicolumn{2}{|c}{\textbf{H1}} & \multicolumn{2}{|c|}{\textbf{L1}} & \multicolumn{2}{|c|}{\textbf{H1}} & \multicolumn{2}{c|}{\textbf{L1}} \\ \hline
Parameter & Mean & Std & Mean & Std & Mean & Std & Mean & Std \\ \hline
$\omega$ & 0.97 & 0.16 & 1.06 & 0.15 & 0.96  & 0.19 & 1.05 & 0.17\\ \hline
$\alpha_{max}$ & 1.41 & 0.12 & 1.58 & 0.17 & 1.40 & 0.12 & 1.57 & 0.17\\ \hline
$\alpha_{0}$ & 0.71 & 0.03  & 0.79 & 0.03 & 0.71 & 0.03 & 0.79 & 0.03\\ \hline
\end{tabular}
\end{table}

\subsubsection{Distinguishing Signal from Noise: Shuffled Data Analysis}
\label{results_noise}
\begin{figure}[h!]
    \centering
    \includegraphics[width=1\textwidth]{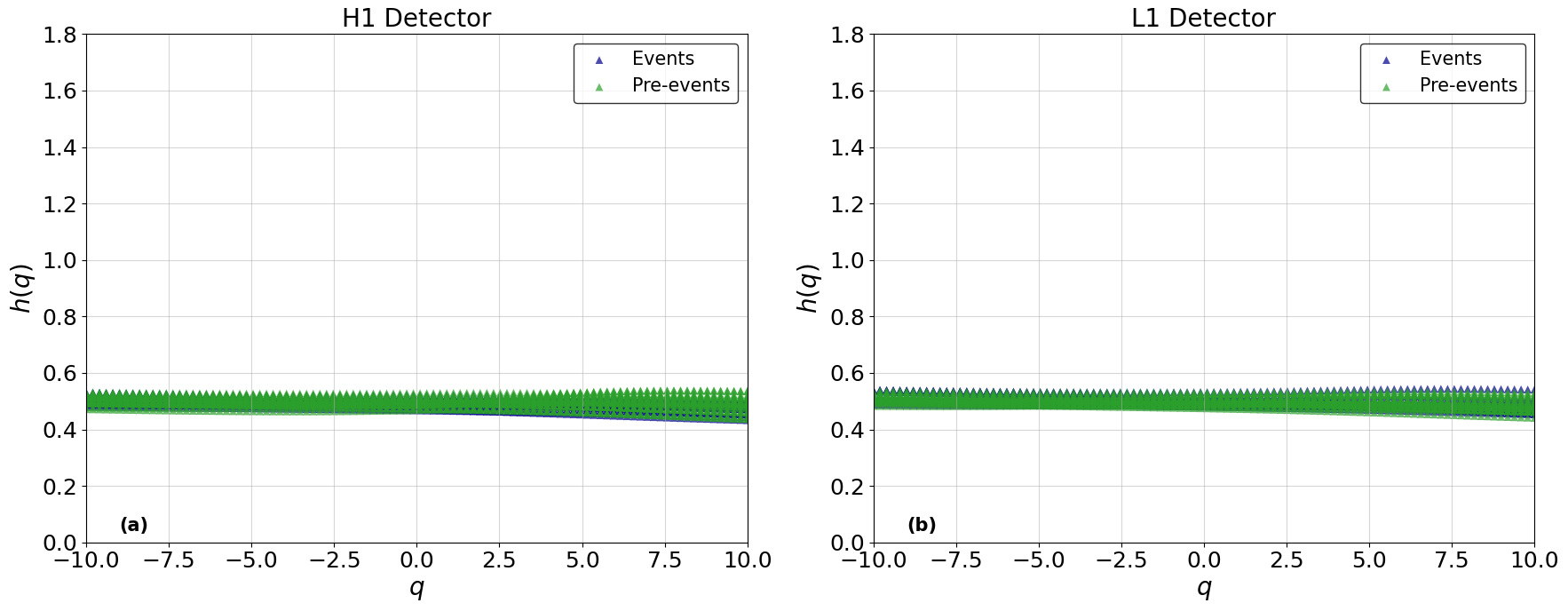}
    \caption{Hurst exponent for shuffled data. The collapse to a constant value confirms temporal correlations are the source of multifractality.}
    \label{fig:shuf_h(q)}
\end{figure}

The shuffle-surrogate test (Figure \ref{fig:shuf_h(q)}) destroys the temporal correlations. As expected, the shuffled data exhibits monofractal behavior ($h(q) \approx 0.5$). This confirms that the multifractality observed in the original series arises from long-range temporal correlations (colored noise) and not from a broad PDF.

\newpage

\subsubsection{Physical Interpretation: Instrumental Noise Dominance and Signal Transience}
\label{sec:interpretation}
Before detailing the physical interpretation of the spectra, it is crucial to clarify the methodological scope of this case study. The primary objective here is not to extract novel astrophysical insights from the gravitational wave event, but rather to benchmark \texttt{MF-toolkit} against a massive, high-frequency empirical dataset. The observation that instrumental noise dominates the multifractal behavior is a key feature of this validation; it explicitly demonstrates the toolkit's capacity to process large-scale data and utilize automated surrogate analysis to successfully disambiguate the structural origins of multifractality. Consequently, this section serves to highlight the software's practical utility as a robust diagnostic metric for instrumental noise characterization. 

Our analysis reveals a significant overlap in the multifractal spectra between the "Event" and "Pre-event" phases. 
Rather than indicating a method failure, this result provides critical insight into the dynamics of the interferometer and explicitly delineates the operational range of multifractal analysis regarding transient signals. To rigorously quantify the observed differences in the multifractal signatures, we conducted a non-parametric statistical analysis using the two-sided Mann-Whitney U test. We focused on the singularity spectrum width ($\Delta\alpha$), which serves as the primary indicator of the degree of multifractality. 

\vspace{2mm}

The statistical comparison between the Hanford (H1) and Livingston (L1) detectors during the actual detection windows revealed a statistically significant difference in their multifractal structures ($p = 0.026$). While both interferometers exhibit broadly similar multifractal features due to their identical technological design, the test successfully captures consistent, subtle deviations. This confirms that the two interferometers possess inherently distinct background noise topologies, likely driven by their different local environmental and instrumental conditions. In contrast, when comparing the ``Event'' data segments to their corresponding background noise ``Pre-event'' segments within the same detector, the statistical tests yielded no significant differences (H1: $p = 0.571$; L1: $p = 0.968$). These high $p$-values explicitly confirm that the observed variations in $\omega$ between an event and its immediate preceding noise are merely descriptive and fall well within the stochastic variance of the baseline noise. Phenomenologically, this implies that the transient gravitational wave signal does not fundamentally alter the macroscopic multifractal scaling behavior of the data block, which remains strictly dominated by the local instrumental noise of each respective detector.

\vspace{2mm}

To interpret this, we must consider the effect of temporal dilution relative to the analysis window. The black hole merger constitutes a high-energy transient typically lasting from fractions of a second to a few seconds within the detector's sensitive band. Conversely, our MFDFA was performed on 32-second windows ($N \approx 5 \times 10^5$ points). Since MFDFA is a statistical measure that averages scaling behavior over the entire observation window, the astrophysical signal—despite its high instantaneous amplitude—is too brief to disrupt the dominant long-range correlation properties of the background noise. Consequently, the multifractal singularity of the event is effectively "diluted" by the persistent dynamics of the instrumental noise that governs the majority of the time series.

Furthermore, the analysis of shuffled data (Subsection \ref{results_noise}) offers a definitive explanation regarding the origin of the observed multifractality. The collapse of the generalized Hurst exponent to $h(q) \approx 0.5$ confirms that the multifractal nature of the LIGO data does not stem from a heavy-tailed probability distribution, such as rare extreme values or isolated high-amplitude glitches. Instead, it arises from nonlinear temporal correlations intrinsic to the detector. This characterizes the LIGO background not as simple Gaussian white noise, nor as a linear fractional Gaussian noise (fGn), but as complex "colored noise." This rich structure likely emerges from the nonlinear coupling between various detector subsystems, including seismic noise, thermal suspension noise, and quantum shot noise, alongside the feedback control loops.

Finally, the consistent discrepancy observed between detectors suggests that MFDFA serves as a distinct fingerprint of the detector state. The L1 instrument exhibits a wider multifractal spectrum width ($\omega$) than H1, independent of the presence of astrophysical events. This implies that MFDFA acts as a sensitive metric for instrumental health, where a wider spectrum in L1 indicates greater heterogeneity in noise time scales or higher intermittency. Therefore, while global MFDFA on long windows serves as a robust tool for characterizing the complexity and non-stationarity of the background noise—crucial for refining noise models in matched filtering algorithms—it requires shorter, localized windows or alternative multiscale approaches to effectively trigger on short-duration transients.

\section{Performance and Availability}
\label{sec:performance}
The computationally intensive steps of MFDFA, particularly the polynomial fitting over thousands of segments, are parallelized using Numba. This provides a significant speed-up on multi-core processors, making the analysis of very long time series (N $> 10^6$) feasible on a standard workstation. The library is designed with a user-friendly, high-level API that automates the complex analysis pipeline.

The results presented in Table \ref{tab:performance_test_revised} confirm that CPU-based parallelization significantly reduces the computational bottleneck associated with the fluctuation function $F_q(s)$. Notably, the performance difference between $N=1$ and $N=2$ cores is negligible ($1.01$ s vs. $0.97$ s). This behavior occurs because the sequential implementation ($N=1$) is already heavily optimized, leveraging low-level vectorized operations that inherently maximize CPU efficiency even on a single Python process. For $N=2$, the overhead introduced by thread management and context switching counteracts any potential parallel gains. However, a clear performance advantage emerges from $N=4$ cores onwards, achieving a $1.84\times$ speed-up. This sub-linear acceleration is expected due to both the thread startup overhead and the inherently sequential portions of the algorithm (such as detrending). Consequently, the toolkit is strategically designed to deliver state-of-the-art performance sequentially, while effectively exploiting the architecture of modern processors when 4 or more cores are allocated.

\begin{table}[htbp]
    \caption{Execution time and acceleration of the MF-toolkit for MFDFA calculation. A comparison with MFDFA package is also given.  The test uses a time series of $N=10^{6}$ points, with scale ranges from $s=200$ to $s=8000$. The results show an acceleration of $1.84\times$ when using 4 cores, which validates the parallel implementation.}
    \label{tab:performance_test_revised}
    \resizebox{\textwidth}{!}{%
    \begin{tabular}{cccc}
    \toprule
        Cores ($N$) & MF-Toolkit Exec. Time ($T_N$) [s] & Speed-Up ($S_N = T_1 / T_N$)  & MFDFA Exec. Time [s] \\
    \midrule
        1 & 1.01 & 1.00  & 1.11 \\
        2 & 0.97 & 1.04  & - \\
        4 & 0.55 & 1.84  & - \\
    \bottomrule
    \end{tabular}%
    }
\end{table}

To complete the performance evaluation of the toolkit, we compared the computational cost of the two automated crossover detection algorithms. Table \ref{tab:crossover_times} details the execution times for both CDV-A and SPIC across varying levels of signal noise. Thanks to the underlying low-level Numba vectorization, both methods exhibit exceptional speed. CDV-A operates almost instantaneously ($\sim 0.04$ s), as its variance-based geometric approach requires minimal overhead. Conversely, SPIC requires roughly an order of magnitude more computation time ($\sim 0.5$ s) due to the $N_p=100$ random iterations demanded by its permutation test. 

It is worth noting that SPIC's execution time slightly increases as the noise level rises. This occurs because the severe stochastic fluctuations force the algorithm to evaluate a more complex mathematical landscape to establish statistical significance. However, the time penalty strictly plateaus around $0.55$ seconds, proving that the method remains highly scalable.

\begin{table}[h!]
    \caption{Computational benchmark of crossover detection algorithms. Mean execution times and standard deviations (in seconds) for the SPIC ($N_p=100$) and CDV-A methods. The benchmark was performed over 100 iterations using a synthetic time series ($N=10^5$ points) subjected to varying levels of additive Gaussian white noise.}
    \label{tab:crossover_times}
    \centering
    \begin{tabular}{ccc}
        \toprule
        Noise Level (\%) & SPIC Exec. Time [s] & CDV-A Exec. Time [s] \\
        \midrule
        0  & $0.37 \pm 0.06$ & $0.039 \pm 0.005$ \\
        5  & $0.50 \pm 0.07$ & $0.042 \pm 0.005$ \\
        10 & $0.51 \pm 0.05$ & $0.044 \pm 0.005$ \\
        20 & $0.55 \pm 0.05$ & $0.045 \pm 0.005$ \\
        30 & $0.55 \pm 0.05$ & $0.046 \pm 0.006$ \\
        \bottomrule
    \end{tabular}
\end{table}

All performance measurements were performed in a controlled environment with an Intel Core i7 processor with 4 cores, 8 logical processors, and 32 GB of RAM, using Python 3.12. Speed-up is calculated relative to the sequential execution time of the MF-toolkit. 

Table \ref{tab:feature_comparison} provides a comparative summary of the features of  \texttt{MF-Toolkit} compared to the MFDFA library \cite{RydinGorjo2022}.

\begin{table}[htbp]
\centering
\caption{Feature comparison between \texttt{MF-toolkit} and the existing \texttt{MFDFA} Python package \cite{RydinGorjo2022}. While fundamental MFDFA computation is effectively handled by prior software, \texttt{MF-toolkit} introduces a fully integrated, automated pipeline that incorporates advanced crossover detection, surrogate data generation for origin analysis, and systematic theoretical validation checks into a cohesive open-source framework.}
\label{tab:feature_comparison}
\begin{tabularx}{\textwidth}{@{}Xcc@{}}
\toprule
\textbf{Analytical Feature} & \textbf{\texttt{MF-toolkit}} & \textbf{\texttt{MFDFA} package} \\ \midrule
Fundamental MFDFA Computation                 & Yes & Yes \\
Parallelized Computation                     & Yes & No  \\
Automated Crossover Detection (CDV-A \& SPIC) & Yes & No  \\
Surrogate Analysis                           & Yes & No  \\
Automated Theoretical Validation             & Yes & No  \\
End-to-End Integrated Pipeline                & Yes & No  \\ \bottomrule
\end{tabularx}
\end{table}

\vspace{2mm}

The \texttt{MF-toolkit} library is open-source and available on GitHub: \url{https://github.com/NahueMendez/mf-toolkit}. The core functionalities depends on \texttt{NumPy}, \texttt{SciPy}, and \texttt{Numba}.

\section{Conclusion}
\label{sec:conclusion}
We have presented \texttt{MF-toolkit}, a new Python library for advanced multifractal analysis. By integrating fully automatic and objective crossover detection algorithms (CDV-A and SPIC), tools for identifying the source of multifractality (IAAFT and random shuffle surrogates), and a robust generator for synthetic test data, \texttt{MF-toolkit} addresses critical limitations of existing MFDFA workflows.

\vspace{2mm}

The practical necessity and rigor of the library were demonstrated through its application to characterize the non-stationary noise in gravitational wave data (LIGO). This application confirmed that the observed multifractal scaling properties are statistically indistinguishable between the event and pre-event detector noise. Critically, by utilizing the built-in surrogate data analysis, the developed software was able to conclusively determine that this multifractality is primarily an artifact of the measuring instrument's noise, rather than a signature of the astrophysical source dynamics. This result underscores the utility of our tool in providing objective constraints on the applicability and interpretation of MFDFA results in complex physical systems.

\vspace{2mm}
Our Python implementation is designed to be efficient, user-friendly, and powerful, facilitating more rigorous and reproducible research into the complex scaling properties of time series.

\vspace{2mm}

Future work will focus on extending and integrating other multifractal analysis techniques, such as the Wavelet Transform Modulus Maxima (WTMM) or Diffusion Entropy Analysis (DEA) method. We hope that \texttt{MF-toolkit} will become a valuable tool for researchers in the broad and interdisciplinary field of complex systems science.

\section{Conflict of Interest Declaration}
The authors declare that they have no known competing financial interests or personal relationships that could have appeared to influence the work reported in this paper. The development of the MF-toolkit library was conducted purely for academic research purposes.

\section{Declaration of Generative AI Use}
During the preparation of this manuscript, the authors utilized a generative AI model (specifically, the Gemini 3.0 model built by Google) exclusively for language editing, improving the fluency, and enhancing the overall clarity and readability of the non-technical prose. The AI was not used to generate or interpret any scientific content, results, methods, or figures, nor was it used to formulate or modify any equations or mathematical expressions. All scientific conclusions and data analyses presented in this paper are the sole original work of the authors.
\newpage

\appendix
\section{Advanced Example of Usage of \texttt{MF-toolkit}}
\label{sec:appendix_code}

Here we provide a Python code snippet illustrating the main workflow of the \texttt{MF-toolkit} library.

\begin{lstlisting}[language=Python, caption=Example workflow with MF-toolkit.]
import numpy as np
from mftoolkit import mfdfa,SPIC
import multiprocessing
import matplotlib.pyplot as plt 

if __name__ == '__main__':
    multiprocessing.freeze_support()
    
    # 1. Generate an example multifractal series (binomial cascade)
    p=0.6
    n_points = 16384
    k = int(np.ceil(np.log2(n_points)))
    N = 2**k
    example_series = np.ones(N)
    for i in range(k):
        num_segments = 2**i
        segment_len = N // (2 * num_segments)
        new_series = np.copy(example_series)
        for j in range(num_segments):
            start_idx = j * 2 * segment_len
                
            # Half of the left: Multiply by p
            left_slice = slice(start_idx, start_idx + segment_len)
            new_series[left_slice] *= p
                
            # Half of the right: Multiply by (1-p)
            right_slice = slice(start_idx + segment_len, start_idx + 2 * segment_len)
            new_series[right_slice] *= (1 - p)
                
        example_series = new_series
            
    
    
    # 2. Define parameters for MFDFA
    
    #.q exponents
    q_values= np.linspace(-5, 5, 100)
    
    #.Scales
    min_scale_val = 200
    max_scale_val = 8000
    num_scales_val = 100 
    scales = np.floor(np.logspace(np.log10(min_scale_val), np.log10(max_scale_val), num_scales_val)).astype(int)
    scales = np.unique(scales) 
    
    #.Order of polynomial for local detrending
    detrend_order=1
    #.Using the whole dataset processing in both directions
    process_both=True
    #.Parallel computation
    num_cores_to_use = 4 
    
    #.Define a subrange for fitting (optional)
    if len(scales) >= 5: # We need some point for defining a subrange
        idx_start_fit = np.where(np.log(scales)>=np.log(300))[0][0]
        idx_end_fit = np.where(np.log(scales)<=np.log(4000))[0][-1]
        if idx_start_fit < idx_end_fit and idx_end_fit < len(scales):
             min_s_fit = scales[idx_start_fit]
             max_s_fit = scales[idx_end_fit]
             custom_scale_range_for_hq = (min_s_fit, max_s_fit)
             print(f"Using custom subrange of scales for fitting h(q): {custom_scale_range_for_hq}")
        else:
            custom_scale_range_for_hq = None
            print("Custom subrange cannot be defined. All valid scales will be used.")
    else:
        custom_scale_range_for_hq = None 
        print("There is no enough scales for defining a subrange. All valid scales will be used")
    
    # 3. Execute MFDFA
    q_valid, h, tau, alpha, f_alpha, Fqs = mfdfa(
        data=example_series,
        q_values=q_values,
        scales=scales,
        order=detrend_order, num_cores=num_cores_to_use,
        segments_from_both_ends=process_both, 
        scale_range_for_hq=custom_scale_range_for_hq   
    )
    
    # 4. Calculate crossovers using SPIC method
    q_to_analyze_idx = np.argmin(np.abs(q_valid - (-5.0))) # Analyze the case of q=-5.0
    q_to_analyze_val = q_valid[q_to_analyze_idx]
    # Obtain Fqs data for that q
    current_F_q_s_row_for_crossover = Fqs[q_to_analyze_idx, :]
    #Again apply valid mask for that q
    valid_F_mask_for_crossover = np.isfinite(current_F_q_s_row_for_crossover) & (current_F_q_s_row_for_crossover > 0)
    
    # Check if there is enough values for finding crossovers
    if np.sum(valid_F_mask_for_crossover) < 2:
        print(f"No enough valid points for finding crossovers with q={q_to_analyze_val:.1f}")
    else:
        best_crossover_indices = SPIC(
            scales[valid_F_mask_for_crossover], 
            current_F_q_s_row_for_crossover[valid_F_mask_for_crossover],
            max_k_to_test=3, 
            num_permutations=50, 
            min_points_per_segment=3, 
            significance_level=0.05,
            n_jobs=4,
            use_numba=True
        )
        print(f"Crossover(s) found for q={q_to_analyze_val:.1f}: {best_crossover_indices}")
        
    # 5. Plot results
    fig, axs = plt.subplots(2, 2, figsize=(14, 10))
    fig.suptitle('MFDFA of a binomial cascade',fontsize=25)
    ax = axs[0, 0]
    ax.tick_params('both',labelsize=20)
    q_index_to_plot = [0, len(q_valid) // 4, len(q_valid) // 2 ,3*len(q_valid) // 4,len(q_valid) - 1]
    for i_q_idx in q_index_to_plot:
        q_val_current = q_valid[i_q_idx]
        if np.any(np.isfinite(Fqs[i_q_idx, :]) & (Fqs[i_q_idx, :] > 0)):
            valid_F_mask = np.isfinite(Fqs[i_q_idx, :]) & (Fqs[i_q_idx, :] > 0)
            ax.loglog(scales[valid_F_mask], Fqs[i_q_idx, valid_F_mask], 'o-', label=f'q={q_val_current:.1f}',zorder=1)

    ax.set_xlabel('Scale($s$)',fontsize=20)
    ax.set_ylabel('$F_q(s)$',fontsize=20)
    #.Custom subrange
    ax.axvspan(min_s_fit, max_s_fit, color='lightgray', alpha=0.5,zorder=0)
    ax.set_xlim(np.min(scales)*0.9,1.1*np.max(scales))
    ax.set_ylim(np.min(Fqs)*0.9,1.1*np.max(Fqs))
    legend = ax.legend(fontsize=15, frameon=True)
    legend.get_frame().set_facecolor('white')      
    legend.get_frame().set_alpha(0.7)             
    legend.get_frame().set_edgecolor('black')
    legend.get_frame().set_linewidth(1.0)
    ax.grid(True, which="both", ls="-", alpha=0.5)
    
    ax = axs[0, 1]
    ax.plot(q_valid, h, 'o-', color='darkblue')
    ax.set_xlabel('$q$',fontsize=20)
    ax.set_ylabel('$h(q)$',fontsize=20)
    ax.grid(True, alpha=0.5)
    ax.tick_params('both',labelsize=20)
    #.tau vs q
    ax = axs[1, 0]
    ax.plot(q_valid, tau, 'o-', color='darkblue')
    ax.set_xlabel('$q$',fontsize=20)
    ax.set_ylabel(r'$\tau(q)$',fontsize=20)
    ax.grid(True, alpha=0.5)
    ax.tick_params('both',labelsize=20)
    
    #.f(alpha) vs alpha
    ax = axs[1, 1]
    valid_spectrum_mask = ~np.isnan(alpha) & ~np.isnan(f_alpha)
    ax.plot(alpha[valid_spectrum_mask], f_alpha[valid_spectrum_mask], 'o-', color='darkblue')
    ax.set_xlabel(r'$\alpha$',fontsize=20)
    ax.set_ylabel(r'$D(\alpha)$',fontsize=20)
    ax.grid(True, alpha=0.5)
    ax.tick_params('both',labelsize=20)
    ax.text(0.95, 0.95, "(d)", transform=ax.transAxes,
            fontsize=15, fontweight='bold', va='top', ha='left')
    plt.tight_layout(rect=[0, 0, 1, 0.96])
    plt.show()

\end{lstlisting}
\bibliographystyle{ieeetr} 
\bibliography{refs} 
\end{document}